\documentclass[aps,prapplied,twocolumn,superscriptaddress]{revtex4-2} 
\usepackage{graphicx}
\usepackage{amsmath}
\usepackage{amsfonts}
\usepackage{amssymb}
\usepackage{float}
\usepackage[colorlinks]{hyperref}
\usepackage{multirow}
\usepackage{bbold}
\usepackage{soul}
\usepackage{makecell}
\usepackage{minibox}
\usepackage{soul}
\usepackage{fancyhdr}
\usepackage{mathtools}
\usepackage{amsthm}
\usepackage{overpic}
\usepackage[pdftex]{pict2e}
\usepackage[dvipsnames]{xcolor}
\usepackage[draft,inline, nomargin]{fixme}

\begin{document}

\title{A prototype reactor-antineutrino detector based on $^6$Li-doped pulse-shaping-discriminating plastic scintillator}

\newcommand{\GT}{George W.\,Woodruff School of Mechanical Engineering, Georgia Institute of Technology, Atlanta, GA, USA}
\newcommand{\IIT}{Department of Physics, Illinois Institute of Technology, Chicago, IL, USA} 
\newcommand{\LLNL}{Nuclear and Chemical Sciences Division, Lawrence Livermore National Laboratory, Livermore, CA, USA}
\newcommand{\NIST}{National Institute of Standards and Technology, Gaithersburg, MD, USA}
\newcommand{\ORNLRx}{Nuclear Nonproliferation Division, Oak Ridge National Laboratory, Oak Ridge, TN, USA}
\newcommand{\ORNLPhys}{Physics Division, Oak Ridge National Laboratory, Oak Ridge, TN, USA} 
\newcommand{\USNA}{Department of Physics, United States Naval Academy, Annapolis, MD, USA} 
\newcommand{\VTCNP}{Center for Neutrino Physics, Virginia Tech, Blacksburg, VA, USA} 
\newcommand{\VTNE}{Department of Mechanical Engineering, Virginia Tech, Blacksburg, VA, USA} 
\newcommand{\Yale}{Wright Laboratory, Department of Physics, Yale University, New Haven, CT, USA} 

\affiliation{\GT}
\affiliation{\IIT}
\affiliation{\LLNL}
\affiliation{\NIST}
\affiliation{\ORNLRx}
\affiliation{\ORNLPhys}
\affiliation{\USNA}
\affiliation{\VTCNP}
\affiliation{\VTNE}
\affiliation{\Yale}

\author{O.~Benevides~Rodrigues}
\affiliation{\IIT}

\author{E.~P.~Bernard}
\affiliation{\LLNL}

\author{N.~S.~Bowden}
\affiliation{\LLNL}

\author{C.~Bravo}
\affiliation{\GT}

\author{R.~Carr}
\affiliation{\USNA}

\author{T.~M.~Classen}
\affiliation{\LLNL}

\author{A.~J.~Conant}
\affiliation{\ORNLRx}

\author{S.~A.~Dazeley}
\affiliation{\LLNL}

\author{M.~T.~Dunbrack}
\affiliation{\GT}

\author{S.~R.~Durham}
\affiliation{\LLNL}

\author{A.~S.~Erickson}
\affiliation{\GT}

\author{A.~Haghighat}
\affiliation{\VTNE}

\author{K.~M.~Heeger}
\affiliation{\Yale}

\author{P.~Huber}
\affiliation{\VTCNP}

\author{A.~Irani}
\affiliation{\IIT}

\author{O.~Kyzylova}
\affiliation{\VTCNP} 

\author{V.~A.~Li}
\email{vali@llnl.gov}
\affiliation{\LLNL}

\author{J.~M.~Link}
\affiliation{\VTCNP}

\author{B.~R.~Littlejohn}
\affiliation{\IIT}

\author{F.~Machado}
\affiliation{\IIT}

\author{M.~P.~Mendenhall}
\affiliation{\LLNL}

\author{H.~P.~Mumm}
\affiliation{\NIST}

\author{J.~Newby}
\affiliation{\ORNLPhys}

\author{C.~Roca}
\affiliation{\LLNL}

\author{J.~Ross}
\affiliation{\VTCNP} 

\author{F.~Sutanto}
\affiliation{\LLNL}

\author{K.~Walkup}
\affiliation{\VTCNP}

\author{J.~Wilhelmi}
\affiliation{\Yale}

\author{X.~Zhang}
\affiliation{\LLNL}

\collaboration{The Mobile Antineutrino Demonstrator Project}

\date{Oct 17, 2025 --- accepted for publication.}

\begin{abstract}
An aboveground 60-kg reactor-antineutrino detector prototype, comprised of a 2-dimensional array of 36 $^{6}$Li-doped pulse shape sensitive plastic scintillator bars, is described. 
Each bar is 50~cm long with a square cross section of 5.5~cm.
Doped with $^{6}$Li at 0.1\% by mass, the  detector is capable of identifying correlated energy depositions for the detection of reactor antineutrinos via the inverse-beta-decay reaction.
Each bar is wrapped with a specular reflector that directs photons towards PMTs mounted at both ends of the bar. 
This paper highlights the construction, key features, and main performance characteristics of the system.
The system, which relies on multiple observables such as PSD, energy, position, and timing, is capable of detecting IBD-like neutron-correlated backgrounds, long-lived decay chains, and cosmogenic isotopes.
\end{abstract}

\maketitle

\section{Reactor Antineutrinos and Safeguards}
The emitted antineutrino energy spectrum and flux carry information on the fuel content and operating conditions of a reactor. 
Antineutrino emission from nuclear reactors is practically unshieldable. 
In principle, reactor operational status and fuel burn-up can be deduced using this signature, as first pointed out in the 1970s~\cite{Borovoi1978}.
Using antineutrino emissions, multiple application experiments were successful in observing reactor status as well as fuel burn-up~\cite{rovno1994, Bowden:2006hu, NUCIFER:2015hdd}.
Reactor-antineutrino detectors may contribute to the nuclear-safeguards landscape~\cite{Bernstein:2019hix, Akindele:2021sbh, Carr2019}. 

Electron antineutrinos emitted by nuclear reactors can be detected through several reactions: electron elastic scattering, coherent elastic neutrino-nucleus scattering (CEvNS), and inverse beta decay (IBD). 
The IBD reaction has been exploited in most reactor-antineutrino experiments, and can be represented as follows:
\begin{equation} \label{eqIBD}
    \bar\nu_e \; + \; {}^{1}\mathrm{H} \; \longrightarrow \; e^+ \; + \; n
\end{equation}
Despite having a 1.8-MeV threshold that rejects $\sim75\%$ of emitted antineutrinos, the IBD reaction (Eq.~\ref{eqIBD}) is advantageous due to its practicality for realizing a detector. The temporal and spatial coincidence of the positron-annihilation (``prompt'') and neutron-capture (``delayed'') signals inside a scalable hydrogenous medium, such as organic scintillator with a neutron-capture dopant, provide strong background suppression.
The delayed coincidence technique and the use of scintillator to observe neutrinos was first achieved in the 1950s by Reines and Cowan in their experiments at the Hanford and Savannah River reactor sites~\cite{Reines1995Nobel}. 

For monitoring applications, reliability and non-intrusiveness are key factors. 
Practical deployment scenarios favor that the detector operate with minimal shielding and overburden. To achieve this, a detector must inherently be capable of excellent background identification and/or rejection. 
The development of stable $^6$Li-doped pulse-shape-discriminating (PSD) plastic scintillators beginning at LLNL in the early 2010s was recognized as having potential for surface-based mobile reactor-antineutrino detectors~\cite{ZAITSEVA2013747}.
Using $^6$Li as a dopant is advantageous since the thermal-neutron capture position can be localized due to the short range of the $^6$Li$(n,t)\alpha$ reaction products,
unlike many commonly used capture agents such as gadolinium, which produce multiple $\gamma$~rays.
Pulse-shape discrimination is a powerful technique based on the difference in time profile of the scintillation light emitted from electronic and nuclear recoils.
To have a detectable rate if deployed at 10s of meters from a $>$100-MW$_{th}$ reactor core, a detector active mass should be in the range of a few hundred kilograms and above. 

Previous efforts to develop mobile reactor-antineutrino detectors that made aboveground  measurements at power reactors are PANDA~\cite{Oguri:2014gta}, miniCHANDLER~\cite{Haghighat:2018mve}, and  VIDARR~\cite{Bridges:2022khb}. 
The PROSPECT experiment also made high statistics aboveground measurements at a research reactor~\cite{PROSPECT:2018dtt,PROSPECT:2018dnc}. 
These experiments used various combinations of PSD, segmentation, and capture agents to implement particle identification and topology selections for background rejection. 
For example, PROSPECT used a $^6$Li-doped PSD liquid scintillator in a two-dimensional (2D) segmented geometry to achieve the several orders of magnitude background suppression needed~\cite{MPM_AAP2018}.  
The applications-focused SANDD program~\cite{Sutanto:2021xpo} 
explored scaled-up volumes of $^6$Li-doped PSD plastic scintillator with properties similar to the liquid scintillator used in PROSPECT.

In an effort to qualify a new $^6$Li-doped PSD plastic material for use in aboveground, mobile reactor antineutrino detection, the Reactor Operations Antineutrino Detection Surface Testbed Rover (ROADSTR) prototype detector has been developed. 
In this work, the ROADSTR detector design is described and its background identification performance is assessed, because this is a critical element to understanding how a full-sized system would perform.
This effort is undertaken as part of the Mobile Antineutrino Demonstrator project, which seeks to advance technical readiness of neutrino-based reactor monitoring concepts by enabling operationally relevant demonstrations using solid-state aboveground antineutrino technologies~\cite{MAD_INMM}.

The paper is structured as follows: first, the hardware of the detector is described, including the scintillator and calibration procedure. 
Then, the key performance characteristics of the system are reviewed, such as the sensitivity to various sources of background events.
The report is concluded with an analysis of the predicted performance in a realistic deployment scenario.

\section{Mobile IBD Detector Hardware}
The operational ROADSTR is built upon a series of successful preceding detectors, which leverage a high-level segmentation with dual-ended PMT readout and PSD. 
In this section, the main components of the system are discussed.  
The system consists of a two-dimensional 
3$\times$3 array of modules each of which is a 2$\times$2 array of scintillator, for a total of 36 bars. PMTs are mounted on both ends of each bar, 72 PMTs in total.

\begin{figure}
\centering
\begin{overpic}[width=1.\linewidth]
{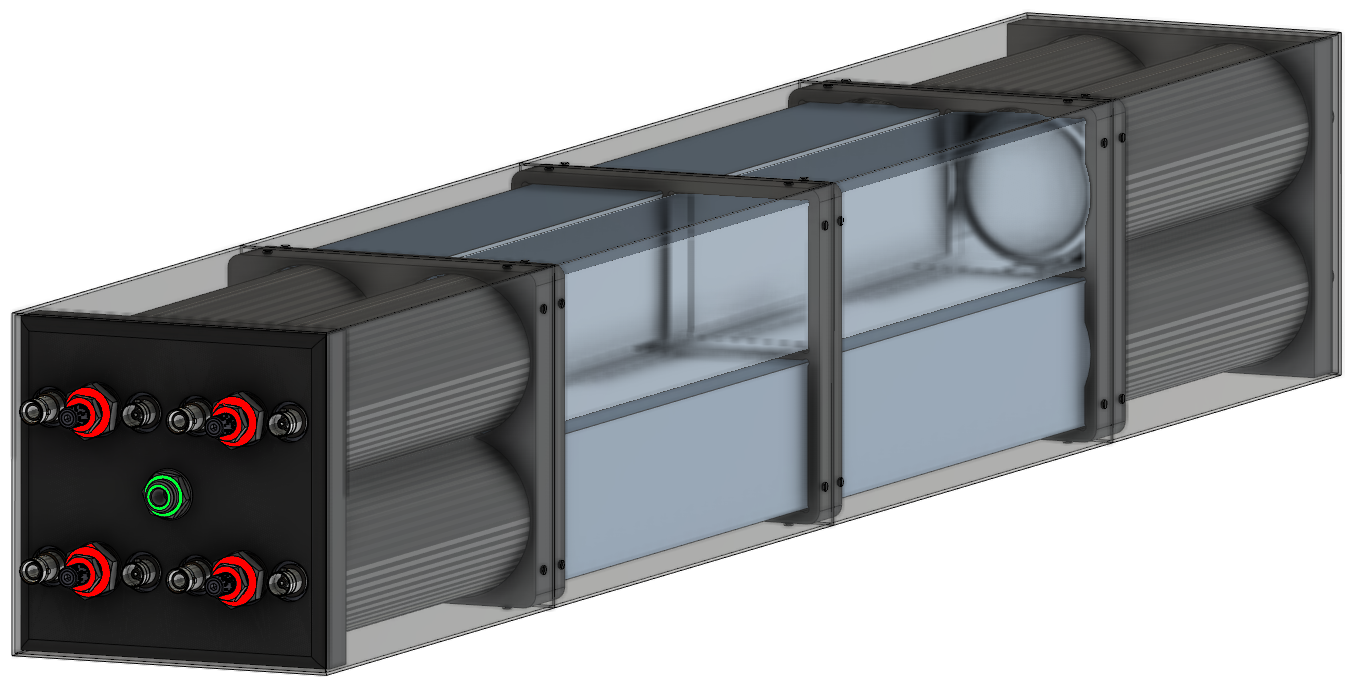}
\put(48,6){\color{red}\vector(0,1){6}}
\put(45,3){\color{black}Scintillator with reflector film}
\put(30, 39){\color{red}\vector(0,-1){13}}
\put(25,40){\color{black}PMT}
\end{overpic}
\caption{A single module contains four 55~mm $\times$ 55~mm $\times$ 500~mm $^6$Li-doped PSD scintillator bars covered with the specular-reflector film (air-coupled to preserve TIR) and eight 2-inch PMTs (coupled with a silicon optical grease). Adjustable spring plungers at the back of the PMTs (shown in red), along with SHV and BNC connectors are visible. The Delrin frames support scintillator bars in the middle and at the ends (4.7~mm and 3.0~mm for inner spacing). The aluminum enclosure and one of the four scintillator bars are made semi-transparent in this rendering. }
\label{fig_module}
\end{figure}
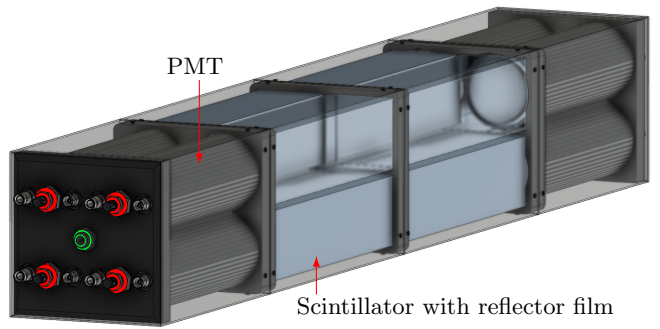

\subsection{Detector Modules}

A modular design was implemented to simplify assembly and to ease detector maintenance. Each module contained 4~scintillator segments and 8~PMTs, which were contained with a light-tight aluminum enclosure, shown in Fig.~\ref{fig_module}.
The aluminum walls were $\sim 0.8$~mm thick.
Although relatively thin, a disadvantage of the aluminium module walls is
that they affect the IBD positron energy reconstruction due to the known $\sim$2~MeV/(g\,/cm$^2$) energy loss.
This was considered to be a reasonable compromise for mechanical reasons in this prototype. 
To support the scintillator bars and PMTs, a series of frames, made of polyoxymethylene (Delrin), are installed at the boundaries of aluminum sub-modules.
Each module is about 1~m long and weighs about 8~kg.
It is designed such that a single person may pick up a module, and assemble multiple modules into a functional detector. 
The modules are positioned on a wire-rope vibration-isolation platform. 
This design could be used in nuclear safeguards to allow rapid deployment and closer installations of IBD detectors in compact and/or difficult to access spaces within a reactor facility.

\begin{figure}
\centering
\includegraphics[width=1.\linewidth]{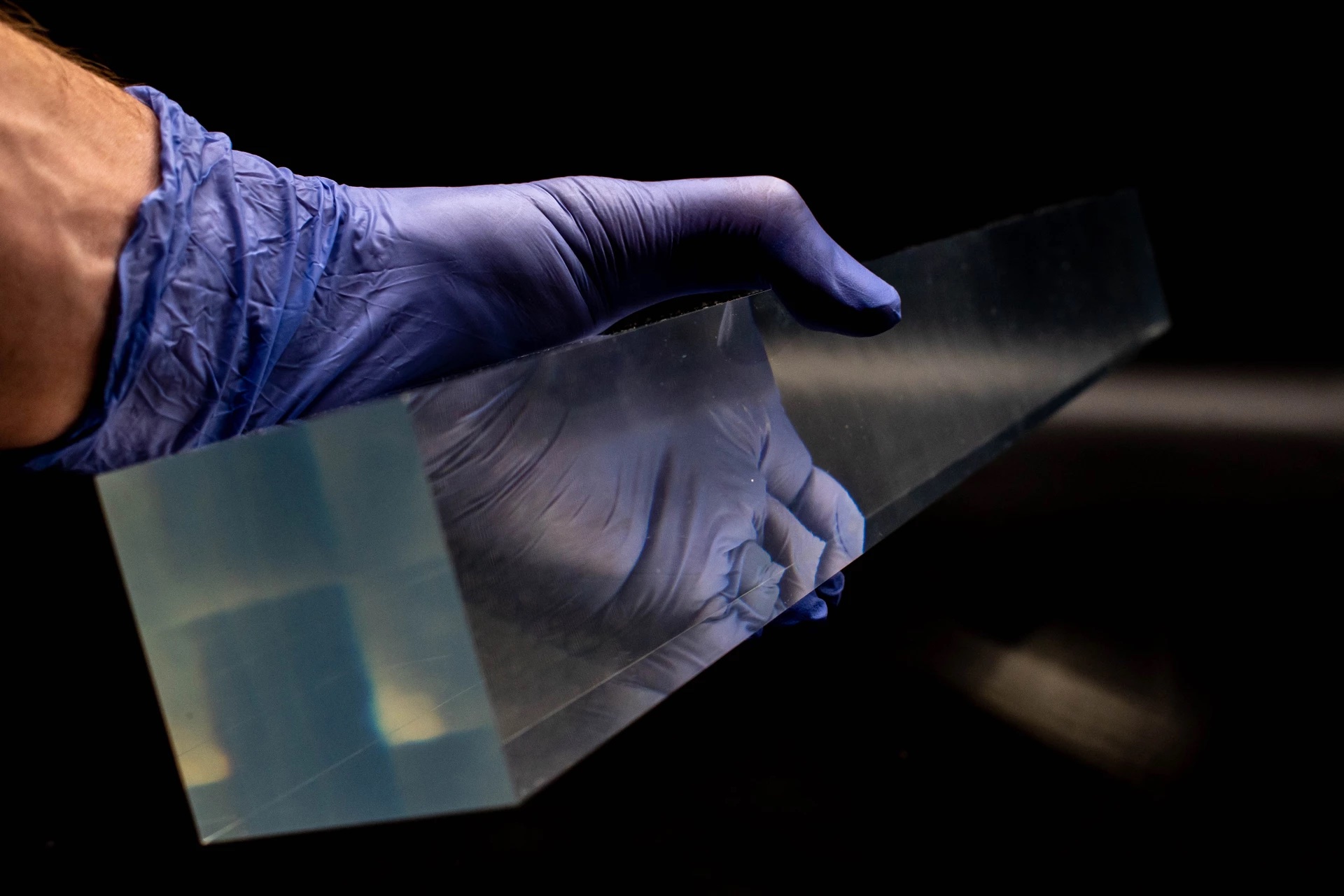}
\caption{A 55~mm $\times$ 55~mm $\times$ 500~mm bar of $^6$Li-doped PSD scintillator.} 
\label{fig_bar_hand}
\end{figure}

\subsection{$^6$Li-doped PSD Plastic Scintillator}

Recent developments in formulation and production have enabled kg-scale castings of $^6$Li-doped PSD plastic scintillators~\cite{Ford:2023ods}. 
Eljen Technology has made further advances in performance and scalability, resulting in material that can be produced at meter-scale lengths with performance suitable for IBD detection~\cite{Roca2024}. 
Designated EJ-299-50, this material has been characterized using collimated $\gamma$-ray sources ($^{22}$Na, $^{54}$Mn, $^{60}$Co, $^{137}$Cs) and a $^{252}$Cf fission source, as described in ~\cite{Roca2024}. For this work, bars of 5.5~cm~$\times$~5.5~cm~$\times$~50~cm were produced by Eljen Technology, shown in Fig.~\ref{fig_bar_hand}, with a  $^6$Li mass fraction of 0.1\%
(1.3$\times 10^{20}$ atoms per cm$^3$), which produces approximately 88\% efficiency for thermal-neutron capture on $^6$Li (the remaining are on hydrogen).

The sides of each bar were covered with a sheet of specular-reflector film (3M DF2000MA)~\cite{3M_ESR_film_2017, 3M_ESR_film_datasheet} to maximize the transport of photons to the PMTs mounted at each end. 
The manufacturer reports the reflectivity of the film is greater than 99\% for the visible light (above 400~nm). The thickness of the film is 38~$\mu$m (66~$\mu$m with an adhesive layer).
There is a small air gap between the reflective non-adhesive side of the film and the bar to allow scintillation light to propagate via total internal reflection (TIR).

\subsection{Photomultiplier-tube Assemblies}

The photosensors selected for this project was the Hamamatsu R7724-100 PMT, packaged in a cylindrical 0.8-mm-thick mu-metal enclosure containing a base with an active voltage divider, to increase output linearity~\cite{HamamatsuCatalog}. The product line for the assembly is H11284-100. 
The outer dimensions of the mu-metal cylinder are 60-mm diameter by 200-mm height.
The assembly is equipped with BNC and SHV connectors to read out the signal and supply the high voltage. 
The PMT is chosen for its relatively high quantum efficiency and fast timing.

From a larger set, a total of 72 PMTs were selected, based on similar gain characteristics.
Separately, PMT pairs for each bar were further chosen on the basis of similar gain at the same voltage. 
This produced a coarse gain-matching of the PMTs. 
A fine tuning of the PMT gain was done using cosmogenic muons, described later in the text. 
PMTs are spring-mounted ($\sim$20-N compression force) and coupled with silicone optical grease (EJ-550, index of refraction of 1.46) to the scintillator.

The dynamic range is set by adjusting the PMT voltages to capture the energy range of the region of interest for reactor-antineutrino detection and most muon tracks. 
PMT voltage values are in the range of 1000--1200~V. 
Although the typical operating voltage of the PMT is 1750~V (max 2000~V) according to the manufacturer, the PMTs are operated at a significantly lower voltage to make sure that the DAQ is not saturated. 

\begin{figure}
\centering
\includegraphics[width=1.\linewidth]{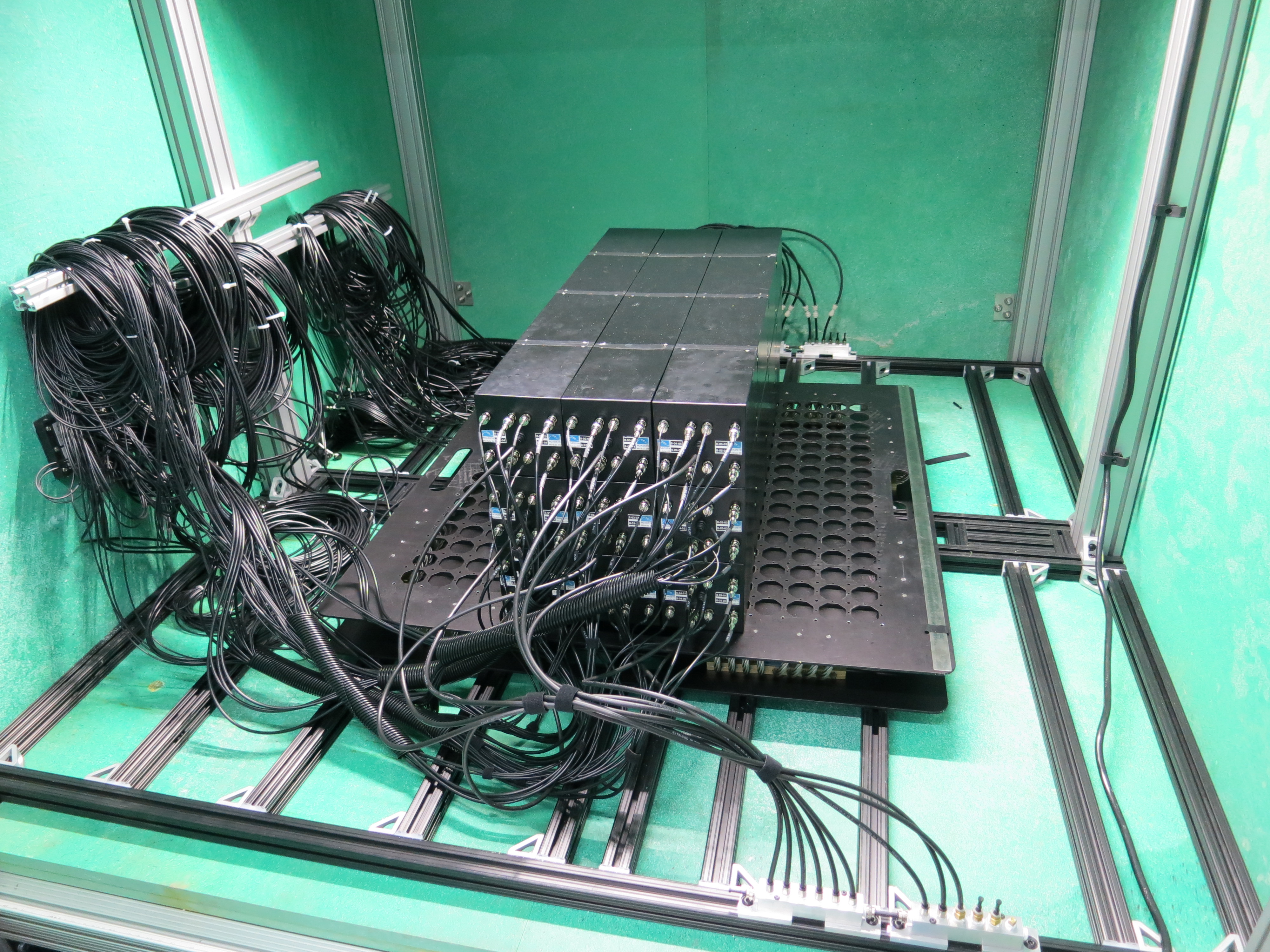}
\caption{A photograph of a the final array of 9 modules, each with four 55~mm $\times$ 55~mm $\times$ 500~mm $^6$Li-doped PSD scintillator bars --- 36 bars with 72 PMTs in total located on a vibration-isolation platform inside 2-inch-thick boron-loaded polyethylene shield. }
\label{fig_3x3array_photo}
\end{figure}

\subsection{Shielding}

An enclosure was constructed for the whole detector to shield it from slow and thermal neutrons. 
The shielding comprises sheets of 2-inch-thick high-density polyethylene sheet (SWX-201HD-G, 5\% boron loaded, 1.08 g/cm$^3$) secured to an aluminum frame, shown in Fig.~\ref{fig_3x3array_photo}. 
The outer dimensions of the shield are 2.03~m $\times$ 1.88~m $\times$ 1.44~m; it weighs 885~kg (1225~kg with the aluminum frame/support structure).
There are double doors on the front and back of the shielding enclosure.
One side has feedthroughs to pass high voltage and signal cables.
The shielding was originally designed to allow expansion of the detector inside if needed, so the inner volume of the shield is therefore much larger than the constructed prototype. 

\subsection{Data Acquisition and Trigger Logic} 

ROADSTR implements a triggering and data acquisition scheme similar to that used by PROSPECT~\cite{PROSPECT:2018dnc}.  
This scheme provides efficient triggering for events with energy depositions distributed across a segmented detector, while maintaining reasonable data collection rates through zero-suppression. 
Data acquisition for all bars is triggered when both of the PMTs on one bar produce a signal 30 analog-to-digital converter (ADC) counts, corresponding to $\mathcal{O}$(10~photoelectrons),  above baseline within a 16-ns (4-sample) coincidence window of each other. 
The formation of this coincidence  is implemented within the firmware of the V1725S waveform digitizer (WFD) and mirrored to a front panel connector as a NIM logic signal. 
That logic signal is distributed to a trigger input on all WFDs via a CAEN V976 logic fan-in fan-out  module. 

On receipt of a trigger signal, the WFD channel for each PMT records a range of [$-$120, 160]\;ns around the point when the waveform rises above the threshold of 15\;ADC counts, up to a maximum waveform length of 300 samples (using the ``Zero Length Encoding'' (ZLE) algorithm implemented in the WFD firmware).
This results in no waveform data being recorded for PMTs with no samples above the ZLE threshold, thus reducing required data rates.
No new triggers are accepted within the 300-sample maximum acquisition window, resulting in a $\leq 1.2$-$\mu$s deadtime for subsequent events.

\section{Performance Characteristics}

Having PSD capability, $^6$Li doping, and segmentation permits the identification 
of unique signatures to test the detector performance. The ready availability of radioactive check sources at a reactor site can not be guaranteed. 
Therefore, various ways are presented to calibrate and characterize the detector without using neutron and $\gamma$-ray radioactive sources.
Fast-neutron-induced backgrounds dominate near the surface; therefore, for testing the IBD performance of the detector,  neutron-capture-correlated events are of special interest, as well as the detector's capability to subtract accidental backgrounds.

\begin{figure}
\centering
\includegraphics[width=1.\linewidth]{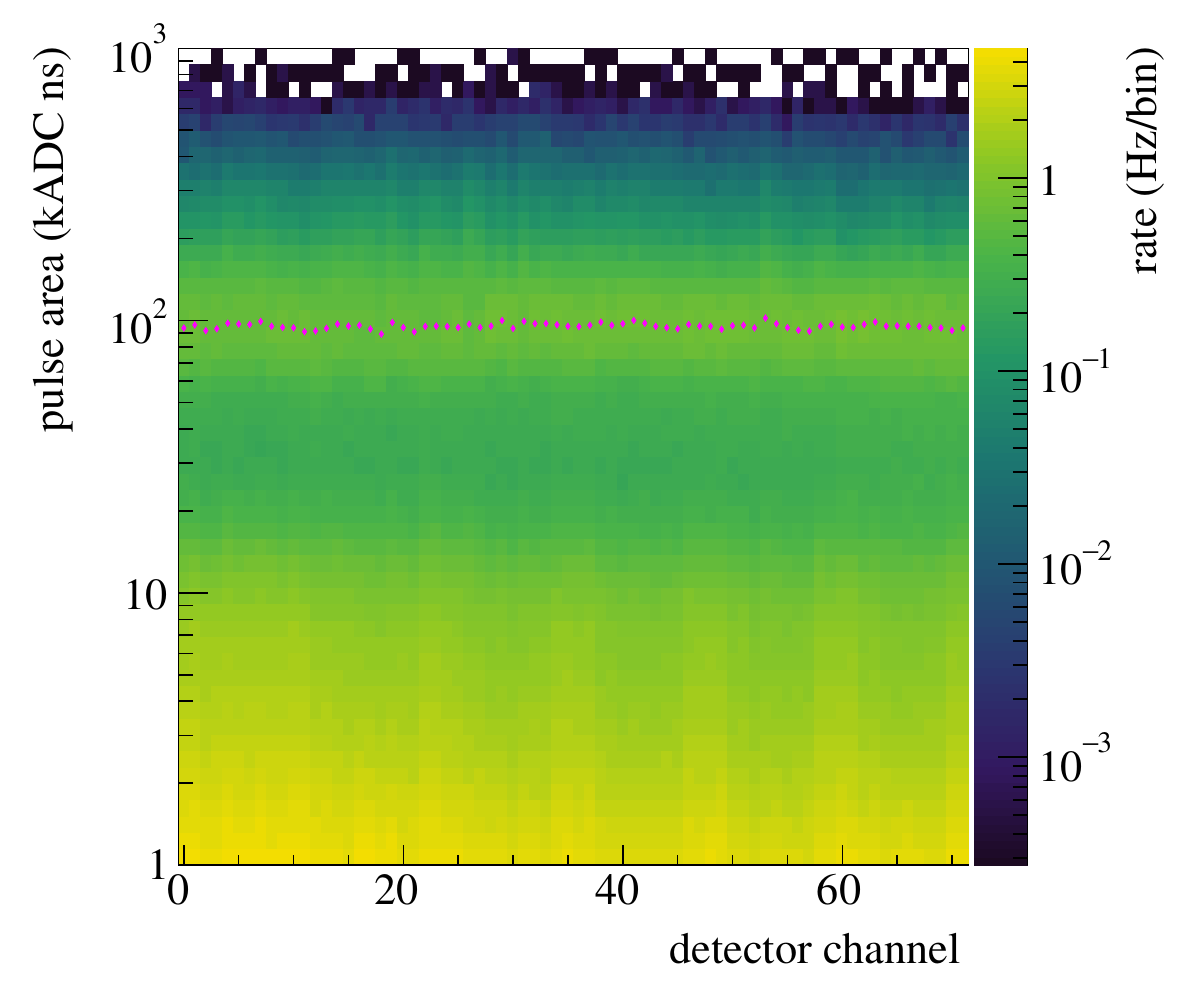}
\caption{
Charge spectra per PMT channel after adjusting gain using muons. As the spectrum has no strong features at higher energy, the gain is adjusted using an arbitrary reference point in each spectrum, indicated by red points, at which 3.3 Hz of events have higher energy.  This is approximately the 10-MeV minimum-ionizing peak in our bar geometry.}
\label{fig_hSpectra_gain}
\end{figure}

\subsection{Energy, Position, and Timing Calibration Using Cosmogenic Muons}

Cosmogenic muons provide an abundant calibration source.
Muons can be selected on the basis that they produce hits in at least two bars.
Since muons are minimum ionizing particles they deposit approximately $\sim$2~MeV per centimeter in organic scintillator. 
They can therefore be used as an environmental high-energy calibration source. Vertical muons deposit roughly 10 MeV per bar ($\sim$5-cm average path length). 
The PMTs were tuned to approximately equal gain based on muon events by equalizing the event rate above a integrated pulse area threshold set sufficiently high that muons predominately contribute (Fig.~\ref{fig_hSpectra_gain}). 
In a detector with dual-ended readout, it is important to note that vertical muons that pass close to one side of a bar will result in a large pulse in one of the two PMTs while the other PMT will have a smaller pulse.
To ensure that the total energy deposited over the full length of each bar can be reconstructed, the gain was set to ensure no saturation of the digitizer was caused by the events occurring close to one side of the bar.

\begin{figure}
\centering
\includegraphics[width=1.\linewidth]{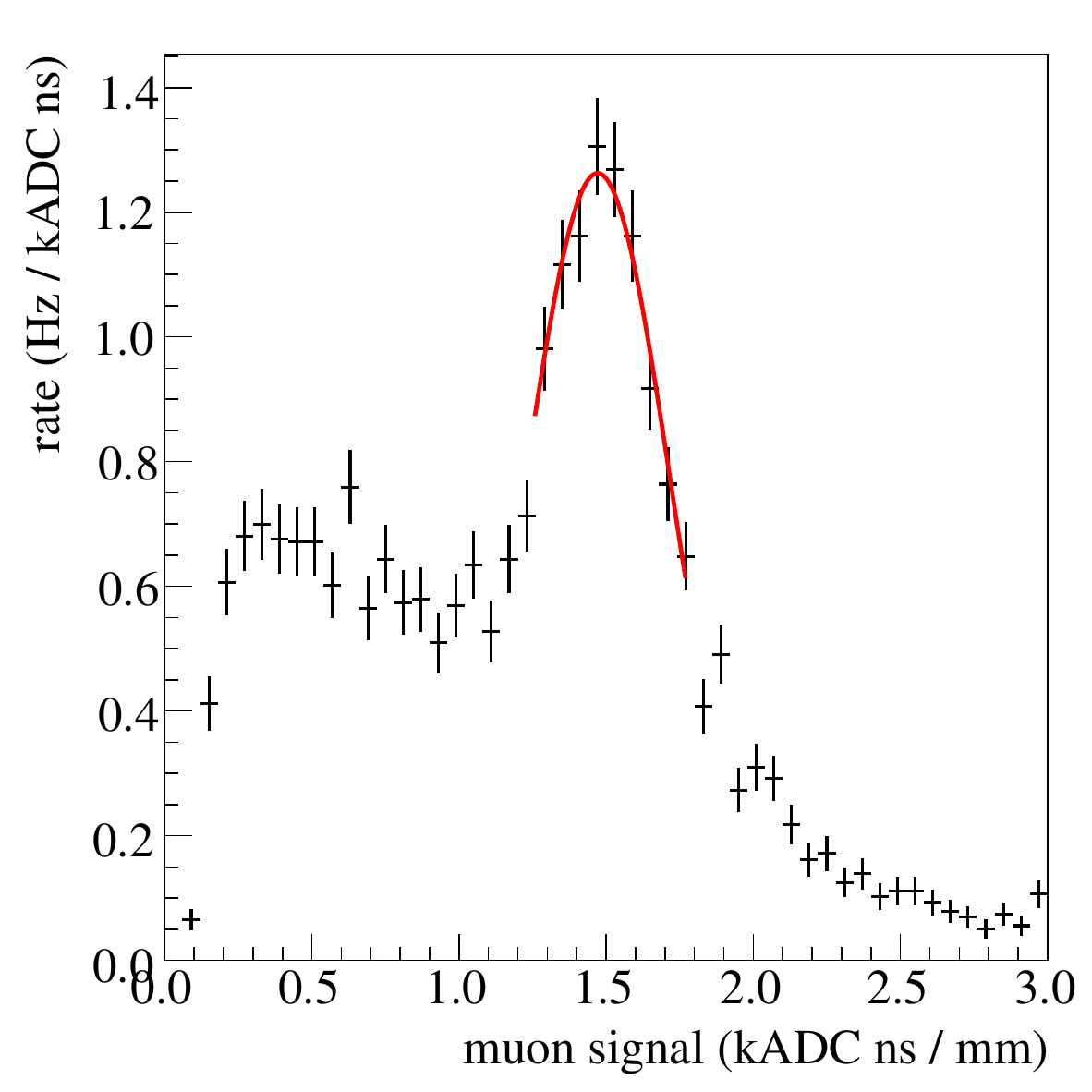}
\caption{A muon minimum ionizing peak per unit of estimated distance for a single representative segment, with a Gaussian fit (shown in red). 
A GEANT4 model was used to match it for a typical muon energy deposition.
} 
\label{fig_muonCali}
\end{figure}

An initial gain match was made within a few percent (Fig.~\ref{fig_hSpectra_gain}).
Software gain corrections were continuously updated from muon data collected during the preceding hour (Fig.~\ref{fig_muonCali}).
This correction accounts for long-term changes in detector response, including the possibility of PMT gain drift, changes in PMT-to-bar coupling efficiency, and gradual accumulation of component chemical precipitate on bar surfaces, which has been observed previously and can alter the light transport efficiency (as described in~\cite{Roca2024}).

After the PMT gains were equalized, 
the ``electron equivalent'' energy per segment is calculated as follows:
\begin{equation}
    E_i = \epsilon_i \sqrt{Q^0_i Q^1_i}
\end{equation}
where $Q^0_i$ and $Q^1_i$ represent the integrated total charge for the two PMTs (0 and 1) for segment $i$ (0--35); the normalization factor $\epsilon_i$ is calculated based on the Gaussian fit to the muon minimum-ionizing peak. 
For multi-segment events the energy is a sum over energies of the individual segments:
\begin{equation}
    E = \sum_i E_i
\end{equation}

To equalize the timing difference from bar-to-bar, a set of muons that hit adjacent bars was selected. 
Using this data set, the average timing offsets from bar to bar were adjusted to equal zero. 
The bar-to-bar timing offsets were stored in a calibration database for subsequent analysis. 

An algorithm was developed to reconstruct the $z$-position of events within each bar. 
Within each bar, a set of muons was selected. For PMTs on either end of each bar, the $\it{average}$ timing difference should equal zero. Under this assumption, each bar offset adjustment is calculated and recorded in a database. 
The relative timing difference to $z$~position is converted using a simplistic constant-speed-per-distance model:
\begin{equation}
    z = \zeta \; \Delta t
\end{equation}
where $\zeta$ is a parameter of the fit.

Both timing and light ratio are used in the position algorithm. 
A position estimate is formed from each independently. 
The two position estimators are combined (weighted average) into the final $z$ position.

\subsection{Pulse-shape Discrimination}

Prior to full detector operation, each bar was characterized. 
For each pedestal-subtracted presmoothed waveform (Gaussian convolution with $\sigma = 4$~ns), 
the pulse-shape parameter $\Pi$ was calculated as a ratio between tail $Q_{tail}$ and total $Q_{total}$ charge, i.e. the area under the curve with integration times determined via a figure of merit optimization: [40,\;400]-ns for tail and [$-$16,\;400]-ns for total charge (the time is relative to the leading edge).
\begin{equation}
    \Pi = \frac{Q_{tail}}{Q_{total}} 
\end{equation}

For combining the PSD from the two PMTs on each end of a bar ($\Pi_0$ and $\Pi_1$) into a single PSD parameter $\Pi$, the number of photoelectrons ($N_0$ and $N_1$) is estimated as seen by each PMT (0 and 1), and the statistically optimal combined PSD parameter is formed:
\begin{equation}
    \Pi = \frac{\Pi_0 N_0 + \Pi_1 N_1}{N_0 + N_1}
\end{equation}

\begin{figure}
\centering
\includegraphics[width=1.\linewidth]{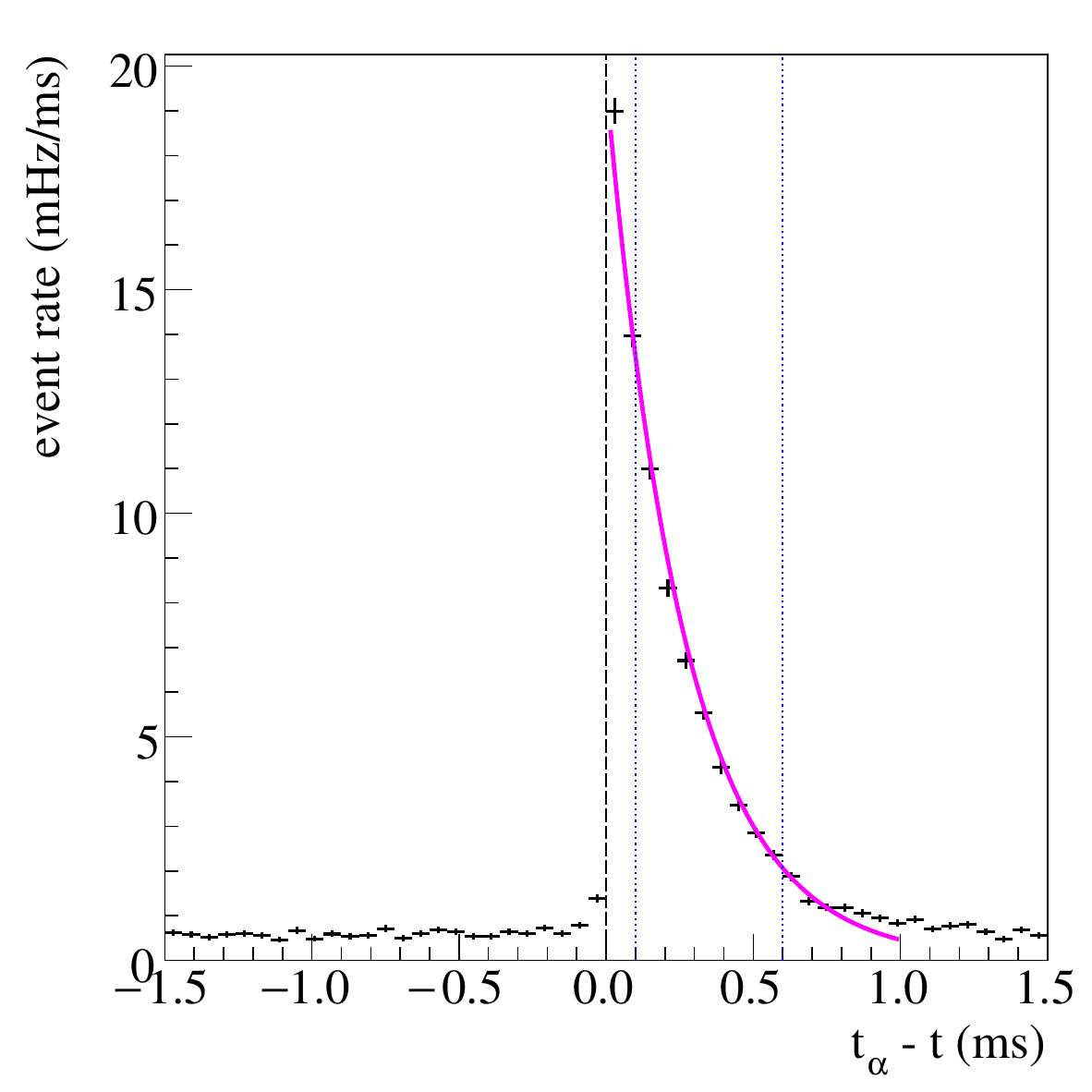}
\includegraphics[width=1.\linewidth]{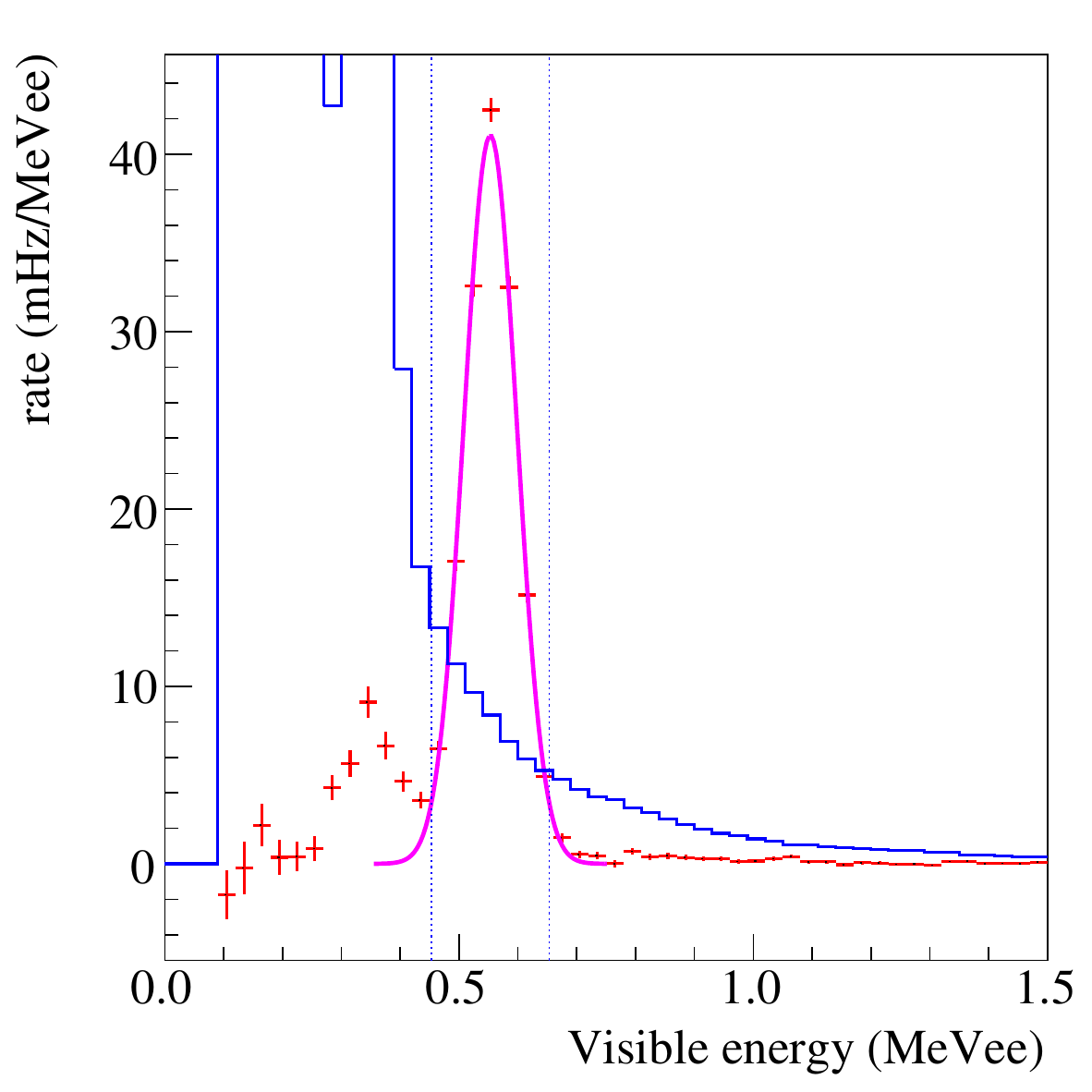}
\caption{The identification of $\beta - \alpha$ coincidences ($\tau$ of 0.237~ms measured) in the decay chain of $^{222}$Rn ($^{214}$Bi$ \to ^{214}$Po$ \to^{210}$Pb).  {\it Top:} 
timing separation between the $\beta$-decay and $\alpha$-decay events.
{\it Bottom: }The rate of $\alpha$~particles  after selection criteria on PSD and timing were applied: the $\sim$0.55-MeV$_{ee}$ peak is due to the heavily quenched scintillator response to 7.833-MeV $\alpha$~particles.
The $\sim$0.35-MeVee peak stems from similarly quenched thermal-neutron captures on $^6$Li, which produce coincident triton and $\alpha$ particles with energy 2.73~MeV
and 2.05~MeV, respectively. 
The blue curve is the accidental backgrounds (mainly from edge clipping muons).}
\label{fig_radon_chain}
\end{figure}

\subsection{Radon-222 $\beta$--$\alpha$ Chain}

Radon atoms forming in earth's crust due to decay of naturally occurring uranium and thorium can diffuse as gas and typically have higher concentration in confined spaces (mines, caves, buildings).
The detector presented in this paper is sensitive to the $\beta - \alpha$ events from the $^{222}$Rn chain (from $^{238}$U series):
$^{214}$Bi$ \to ^{214}$Po$ \to^{210}$Pb --- with the measured timing constant $\tau = 0.237\;\mu$s for the polonium decay. The 7.833-MeV $\alpha$~decay (Bi--Po) provides a calibration constraint on scintillator heavy-particle quenching (0.55~MeV of observed energy), as shown in Fig.~\ref{fig_radon_chain}.

\subsection{Singles}
\begin{figure}
    \centering
    \includegraphics[width=1.\linewidth]{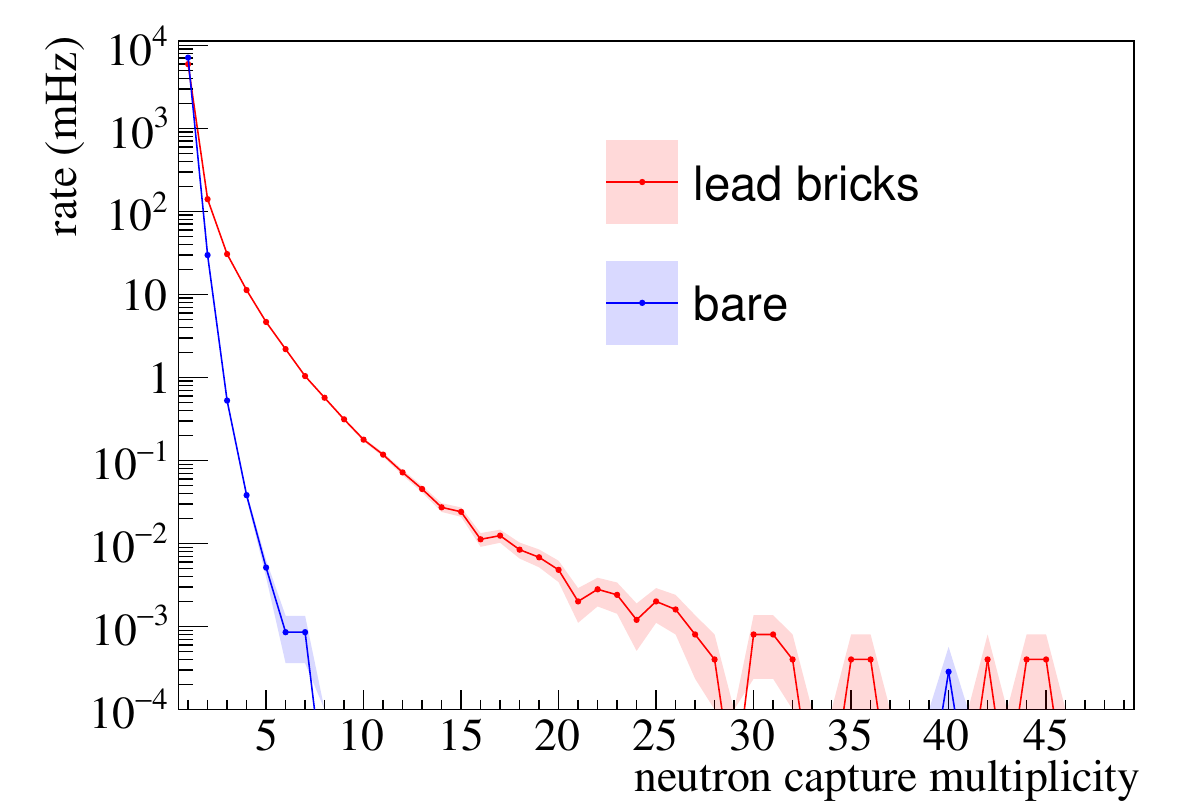}
\caption{
Multiplicity of thermal-neutron-capture candidates within a 100-$\mu$s window. The $y$ axis  is the event rate for the entire detector in two configurations: ``bare'' --- without any additional shielding surrounding the detector, as shown in Fig.~\ref{fig_3x3array_photo}; and ``lead bricks'' --- with three $\sim 5 \times 10 \times 20 \; \mathrm{cm}^3$  lead bricks positioned on top of the detector. The ``lead brick'' induced increase in correlated neutrons is due to the neutrons being knocked out by energetic cosmogenic muons passing through the lead bricks (neutron spallation).}
    \label{fig_hMult}
\end{figure}

Using segmentation, an effect of self-shielding can be observed when scintillator segments in the  outer layers have a higher singles rates, i.e. individual occurrences of energy depositions.
The ROADSTR instrument was used to investigate how accidental and neutron-correlated backgrounds vary with different shielding configurations and in locations with varied overburden. An example of how proximity to high-Z material can result in a large increase in events with high neutron-capture multiplicity is shown in Fig.~\ref{fig_hMult} (thermalization of spallation neutrons on a 100-$\mu$s timescale).
This clearly indicates that ROADSTR is able to measure neutron-capture correlations. A detailed characterization of the neutron response of ROADSTR and the ability of simulations to reproduce measurements under varied shielding conditions is under preparation. 

\subsection{Boron-12 beta decay}

\begin{figure*}
    \centering
    \includegraphics[width=.32\linewidth]{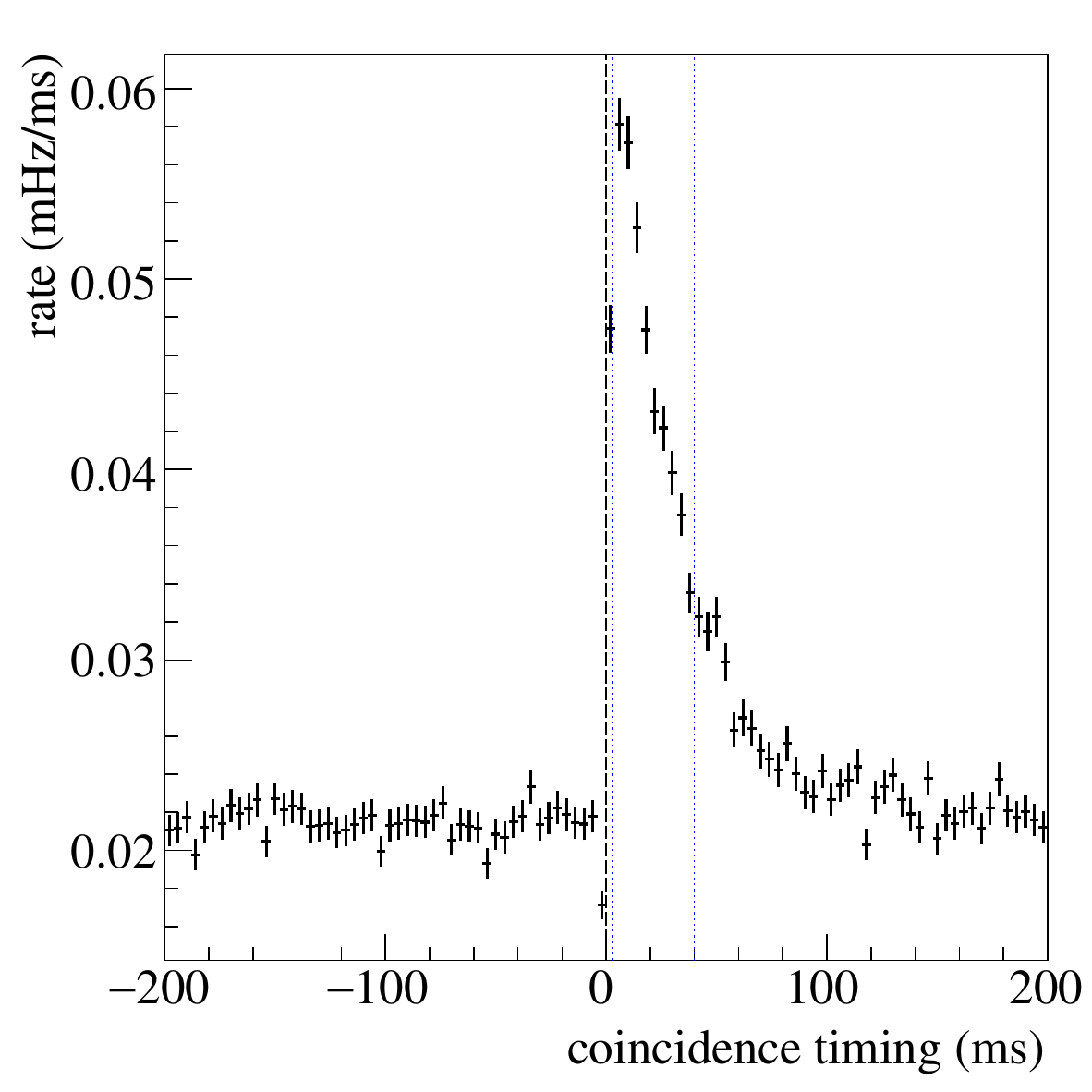}
    \includegraphics[width=.32\linewidth]{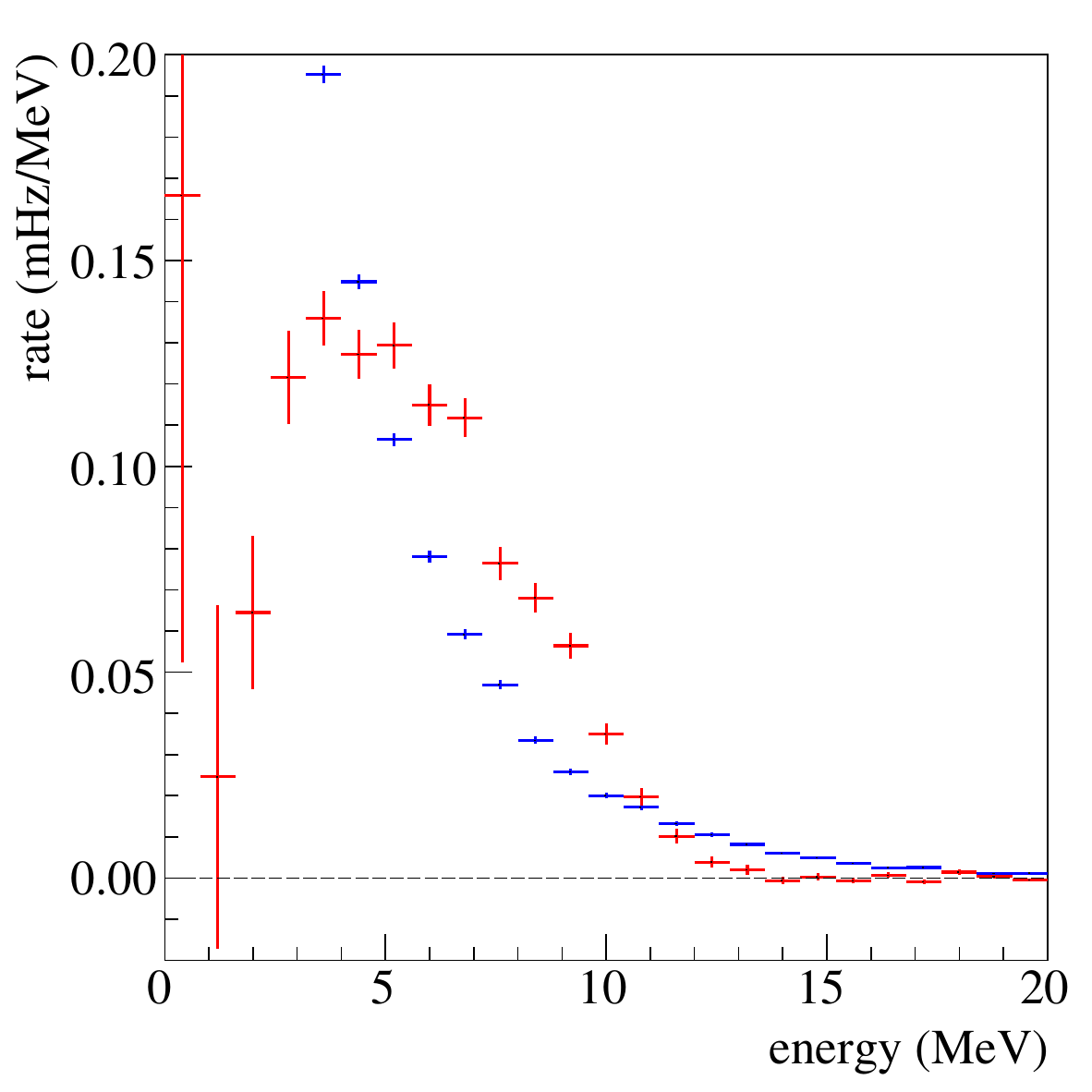}
    \includegraphics[width=.32\linewidth]{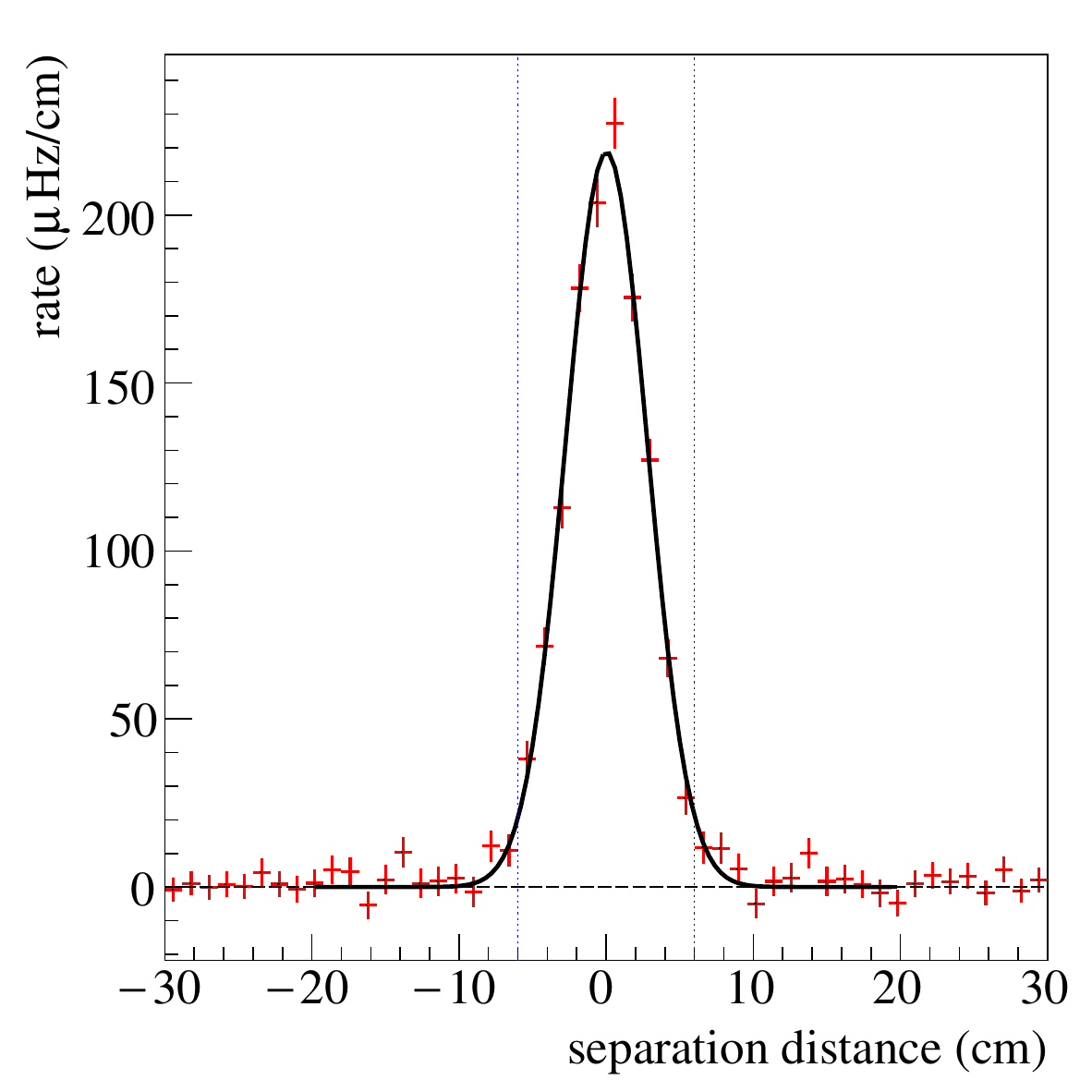}
\caption{Timing, energy, and position distributions for the tagging on prompt isolated proton recoils, that catches $\mathrm{n} + ^{12}\mathrm{C} \to $ $^{12}$B $ + \mathrm{p}$ reaction followed by energetic $^{12}$B $\beta$ decay (13.37-MeV endpoint, lifetime $\sim 30$ ms). 
The red is the energy spectrum of electrons from the $^{12}$B decay; blue --- accidental backgrounds mainly from edge clipping muon events. 
{\it Right:} Tight position correlation in $^{12}$B events (proton recoil observed at same location as subsequent beta decay) helps isolate from backgrounds and provides indicator of position reconstruction accuracy ($\sigma$= 2.85~cm  between proton and beta track). The two vertical dashed lines indicate the distance cuts applied to generate the other two plots shown in this figure, i.e. $[-6\,\mathrm{cm}, 6\,\mathrm{cm}]$.}
    \label{fig_B12_t_E_pos}
\end{figure*}

ROADSTR is able to observe the $\beta$~decay of  $^{12}$B 
cosmogenically produced within the scintillator via $\mathrm{n} + \mathrm{^{12}C} \to \mathrm{^{12}B} + \mathrm{p}$. 
These events provide an additional source for calibrations of timing, position, and energy scale.  
Electrons from the $\beta$ decay of $^{12}$B (lifetime $\sim$30~ms) have a mean energy of $\sim$6.4~MeV with an endpoint of 13.37~MeV.
This provides a high-energy calibration and timing reference (Fig.~\ref{fig_B12_t_E_pos}).
For this reaction, the prompt signal is a nuclear recoil (proton; between 0.7 and 10~MeVee; PSD region as highlighted in Fig.~\ref{fig_IBDlike_XY_Z_t}; single-segment event) and the delayed signal is an electronic recoil (electron).
In addition, the tight position correlation in $^{12}$B events (proton recoil occurring at the same location as subsequent $\beta$ decay) 
provides a direct measure of the position reconstruction  accuracy along the bar length ($\sigma = 2.85$~cm between proton and beta, Fig.~\ref{fig_B12_t_E_pos}).
The ability to pick out this unique reaction is demonstration that this relatively small detector with no overburden can find such reactions (mHz)
which are many orders of magnitude below the background rate (kHz), similar to what is required to identify IBD interactions --- ``finding a needle in a haystack''.

\subsection{IBD-like Event Selection}

The performance of this detector near a nuclear reactor can be projected.
The response of the ROADSTR system to reactor antineutrino IBD interactions was modelled using a simulation toolkit based on GEANT4~\cite{geant4} and  ROOT~\cite{Brun:1997pa}. 
This incorporates detector performance parameters describing quantities like energy resolution, position resolution, and PSD response derived from the measurements presented above and other characterization of EJ-299-50~\cite{Roca2024}. 
Within the simulation, an IBD interaction position is uniformly sampled from within all hydrogenous material in the system, weighted for hydrogen density and $1/r^2$ for an assumed standoff distance of $25$~m from an antineutrino source equivalent to a $2900$-MW$_\mathrm{th}$ reactor. 
The antineutrino energy is sampled from a distribution obtained using a parameterization of fission isotope reactor fluxes~\cite{Huber:2011wv} for highly-enriched $^{235}$U fuel,
    multiplied by the IBD interaction cross-section.
The initial momenta of the IBD positron and neutron reaction products are sampled from distributions that describe the kinematics for the IBD interaction appropriate for the simulated antineutrino energy and momentum along the source-interaction point direction vector.
The reaction products are propagated by GEANT4 until the IBD neutron is captured or leaves the detector volume. 
Ionizing energy depositions in the scintillator are recorded and processed through an approximate detector response model, into the same format as observable data for analysis through the same code.

\begin{figure*}
\centering
\includegraphics[width=.36\linewidth]{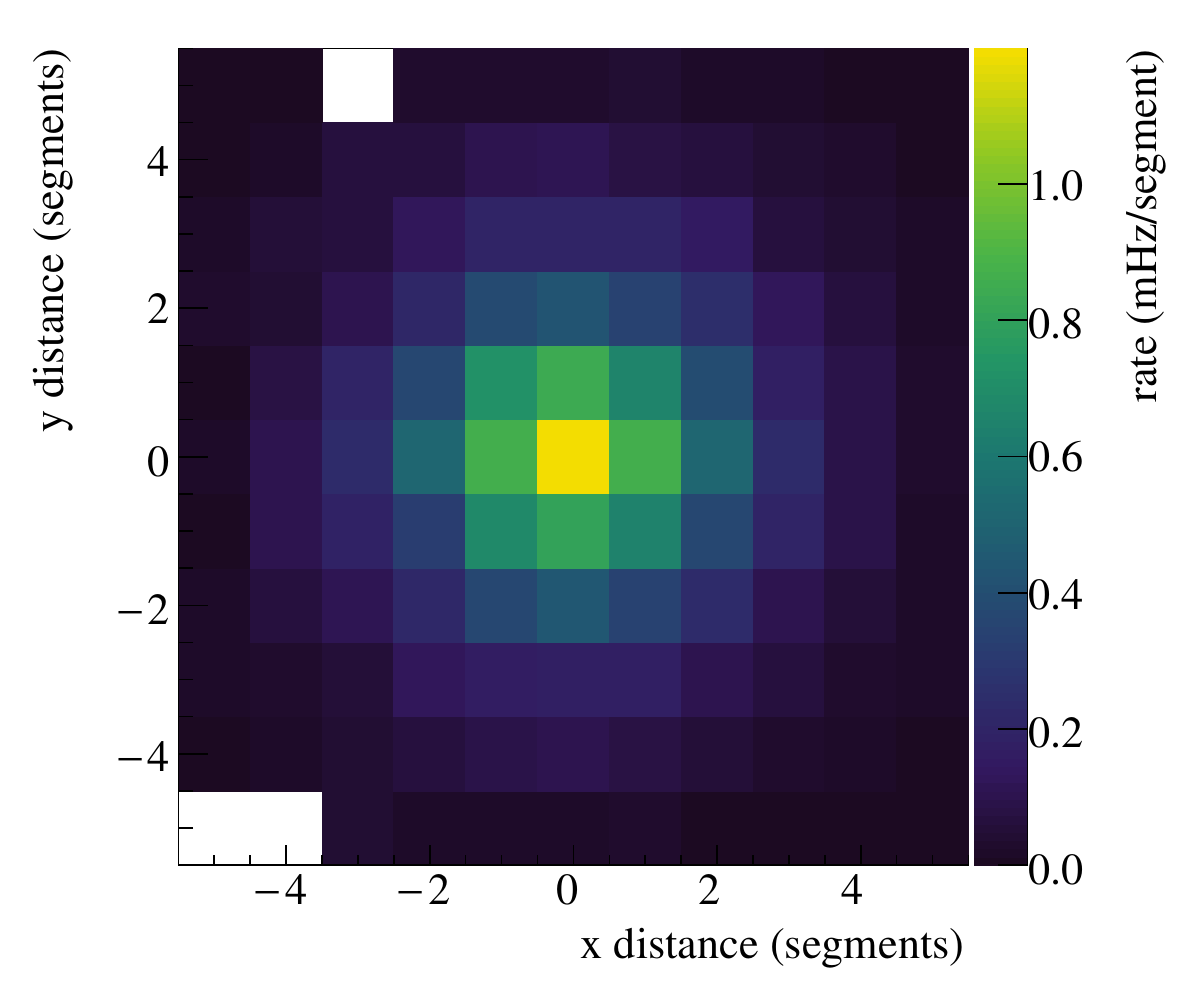}
\includegraphics[width=.30\linewidth]{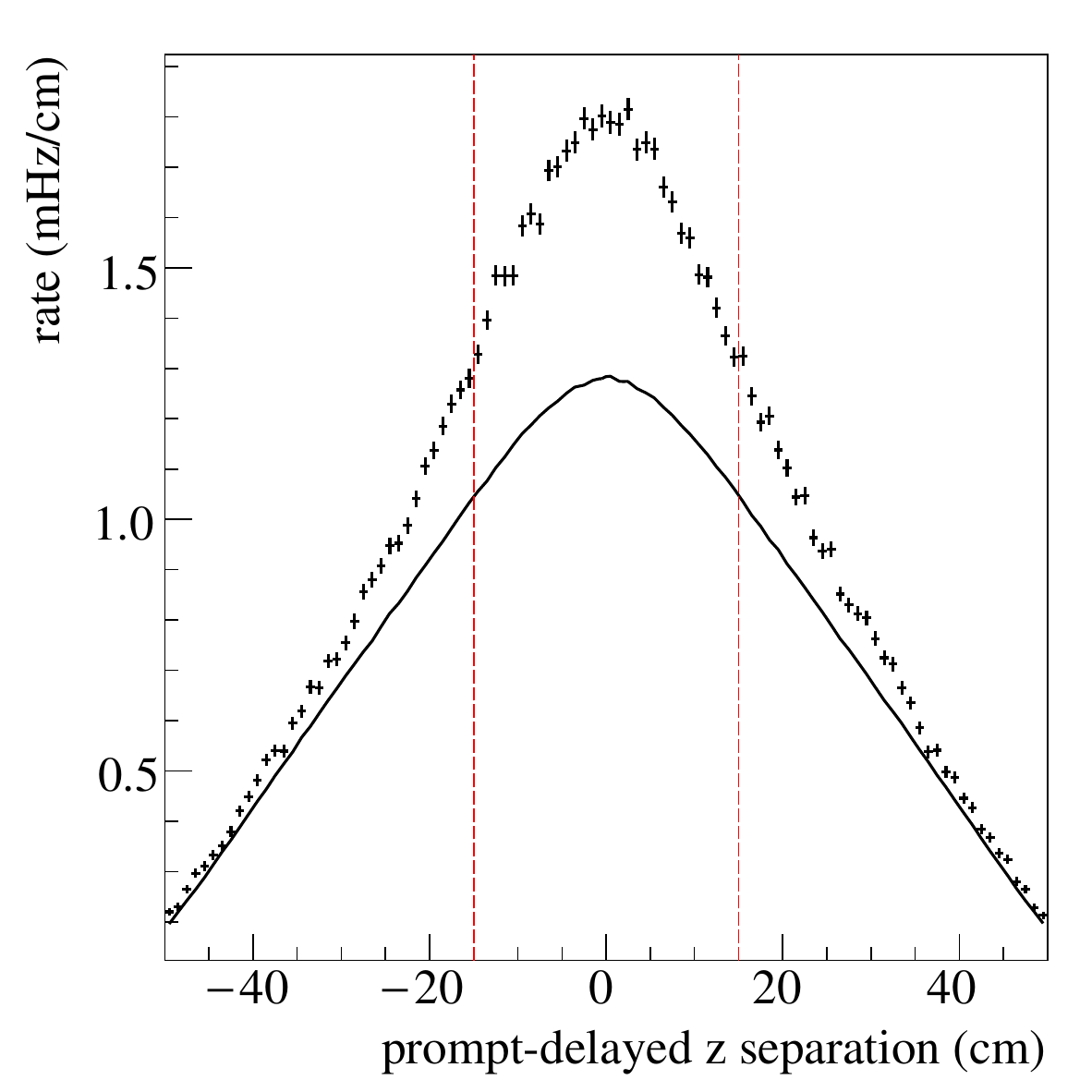}
\includegraphics[width=.30\linewidth]{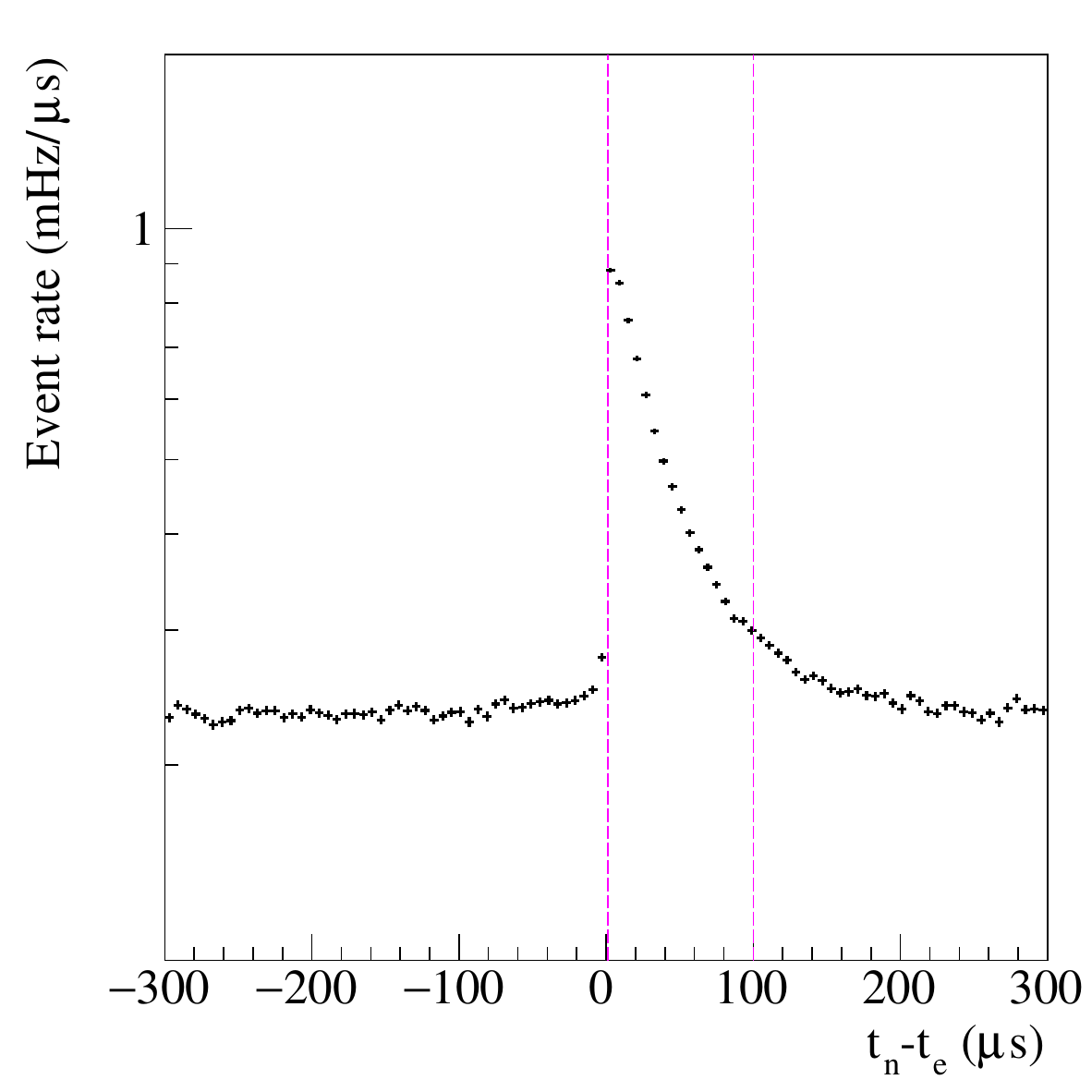} \newline
\includegraphics[width=.36\linewidth]{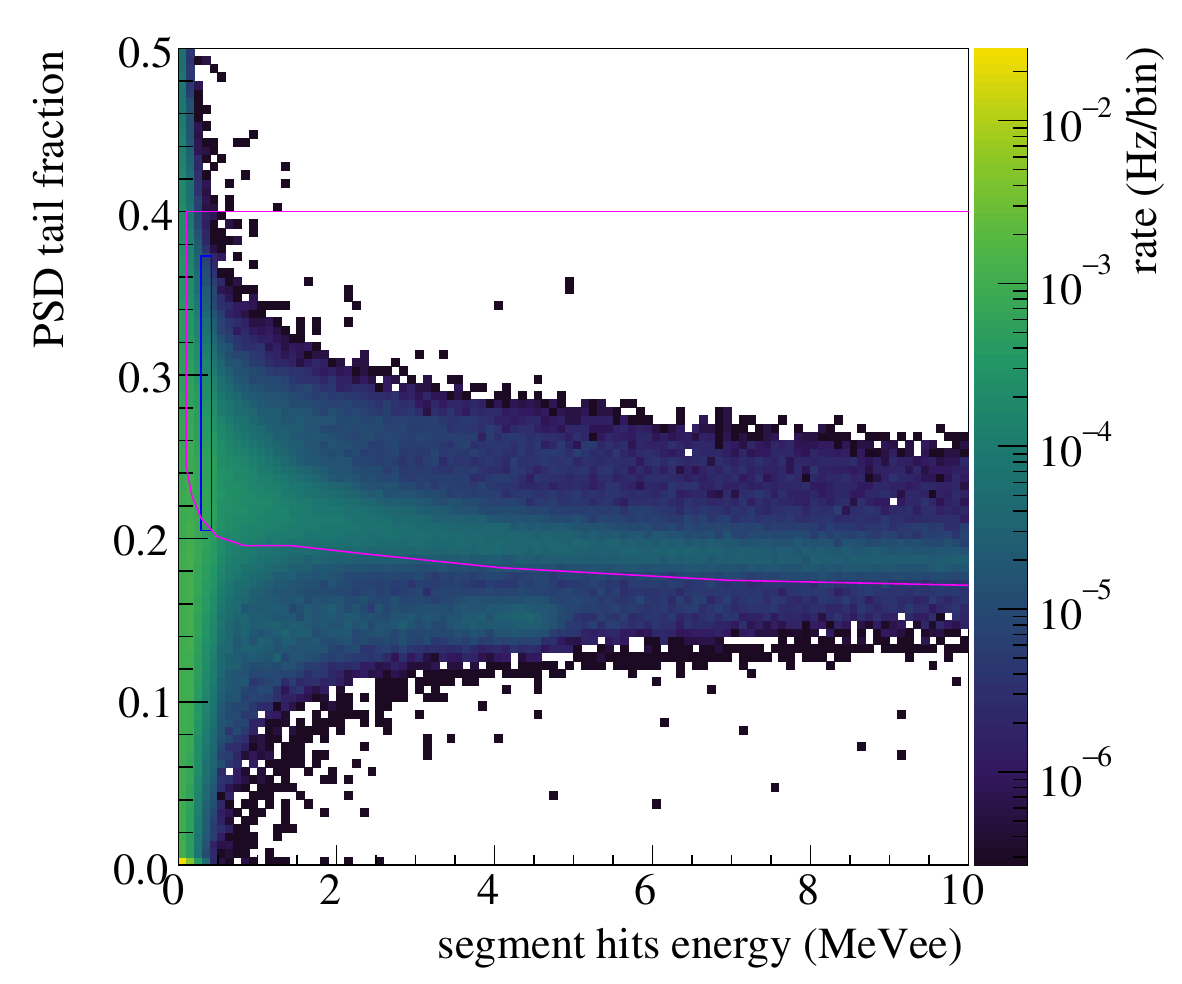}
\includegraphics[width=.36\linewidth]{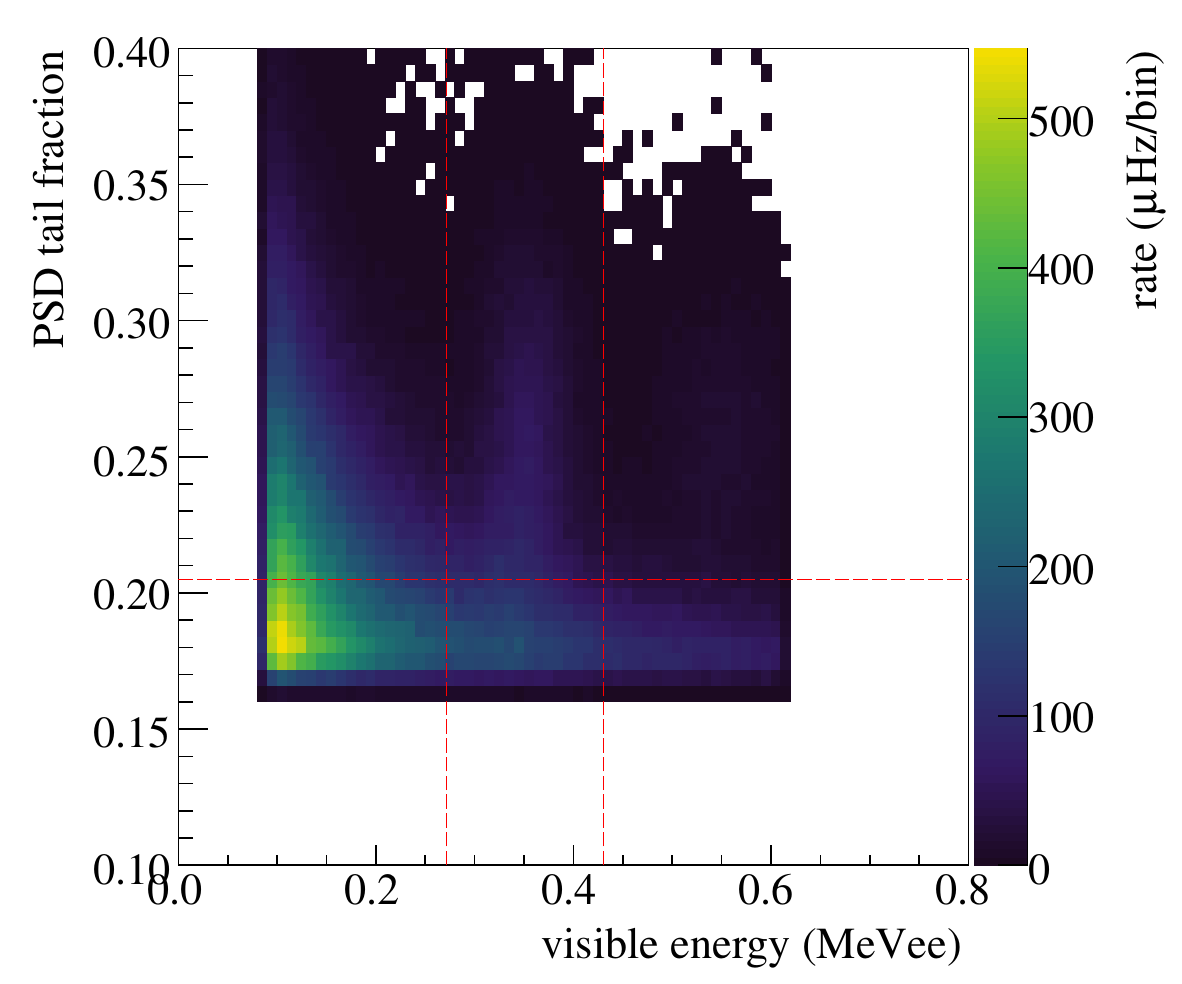}
\caption{IBD-like event selection. (Top row) Position (in $x$-$y$ and $z$) and timing distributions of the delayed signal with respect to the prompt. The cuts on timing and position are indicated by the dashed vertical lines.
(Bottom row) Pulse-shape parameter as a function of energy for prompt (left) and delayed events (right). A cut region imposed on the prompt is outside of the blue and magenta regions.}
\label{fig_IBDlike_XY_Z_t}
\end{figure*}

The following separation and PSD cuts were applied to select IBD-like (neutron-correlated) events:
\begin{enumerate}
    \item Prompt-delayed timing separation --- $1.2\;\mu\mathrm{s} < t_n - t_e < 100\;\mu$s (for all candidate pairs passing cuts in  $1.0\ \mathrm{MeV} \le E_\mathrm{prompt} \le 7.5$~MeV), where $t_e$ is the timing for the prompt and $t_n$ --- for the delayed event.
    \item Prompt-delayed position separation --- the cut is 15~cm in $z$ (along the bar) in the same or the 8 adjacent segments. 
    \item Prompt PSD cut to reject fast neutrons ---
    for multi-segment prompt events (in the IBD-like event selection), the segment with the highest number of sigma deviation from the gamma band is selected (using an estimate for the energy-dependent PSD resolution) --- $\pm 2 \sigma$ about the mean value of the electronic-recoil PSD band (Gaussian fit).
    \item Delayed PSD cut to select neutron captures --- $\pm 2 \sigma$ about the mean value in the PSD ``capture-island'' and $\pm 3 \sigma$ about the mean value in energy (Gaussian fits in PSD and energy to the ``capture-island'').
    \item Vetoes following events identified as muons, fast neutrons, or neutron captures.
\end{enumerate}
The selections and cut values, represented in Fig.~\ref{fig_IBDlike_XY_Z_t}, are similar to those used by previous segmented detectors. Separation cut values are chosen to select regions where correlated signals are predominant over random backgrounds. Particle-identification cut values are similarly chosen to balance desired particle type efficiency with background rejection, considering energy and PSD resolution. Veto values determined by PROSPECT were used since it was demonstrated that those values had reached a point of diminishing return in background rejection~\cite{PROSPECT:2020sxr}, while introducing minor deadtime. Similar or better veto rejection performance can be expected in ROADSTR, since neutron production, transport, and capture timescales are similar, while the smaller size and thus background flux in ROADSTR will yield lower deadtime. 
A full parameter scan of the various cut values has not been performed since an experimental IBD data set is not currently available to optimize against. 

\begin{figure*}
    \centering
    \begin{tabular}{ccc}
        (a) separation + PSD cuts & (b) + energy topology & (c) + geometry topology\\ 
        \includegraphics[width=0.32\textwidth]{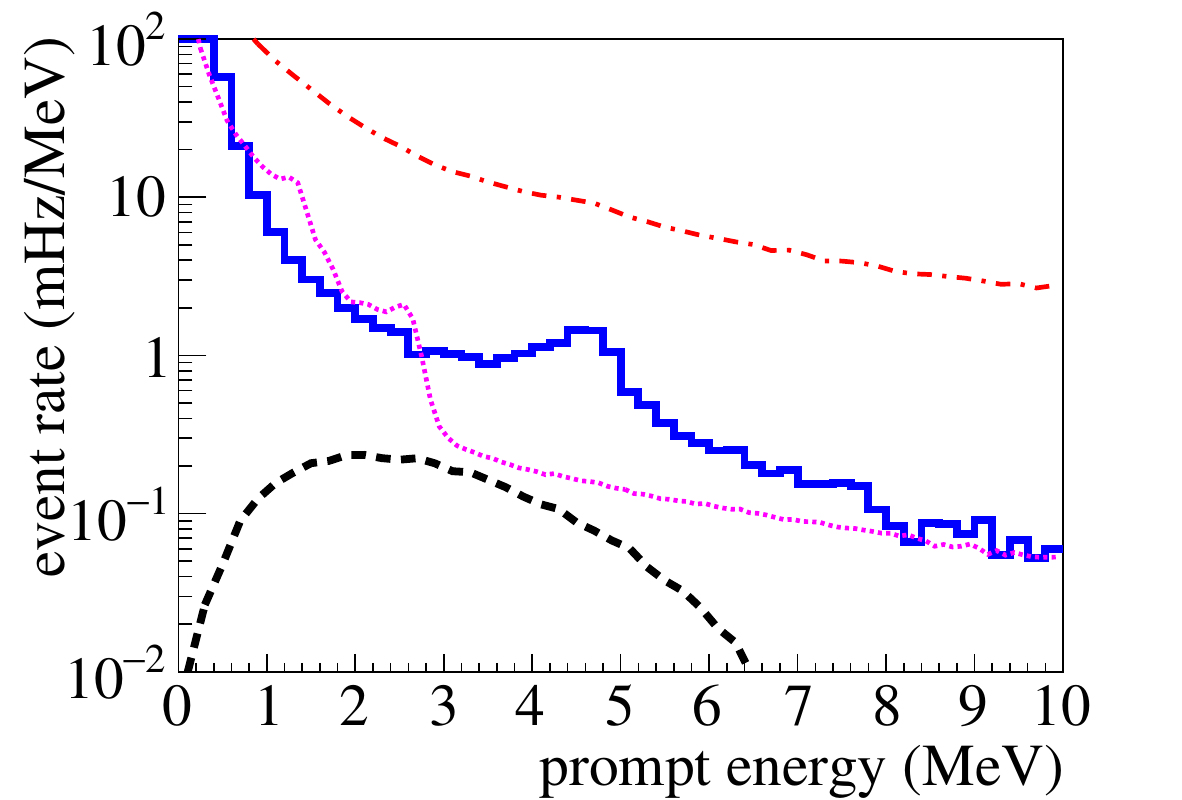} &
        \includegraphics[width=0.32\textwidth]{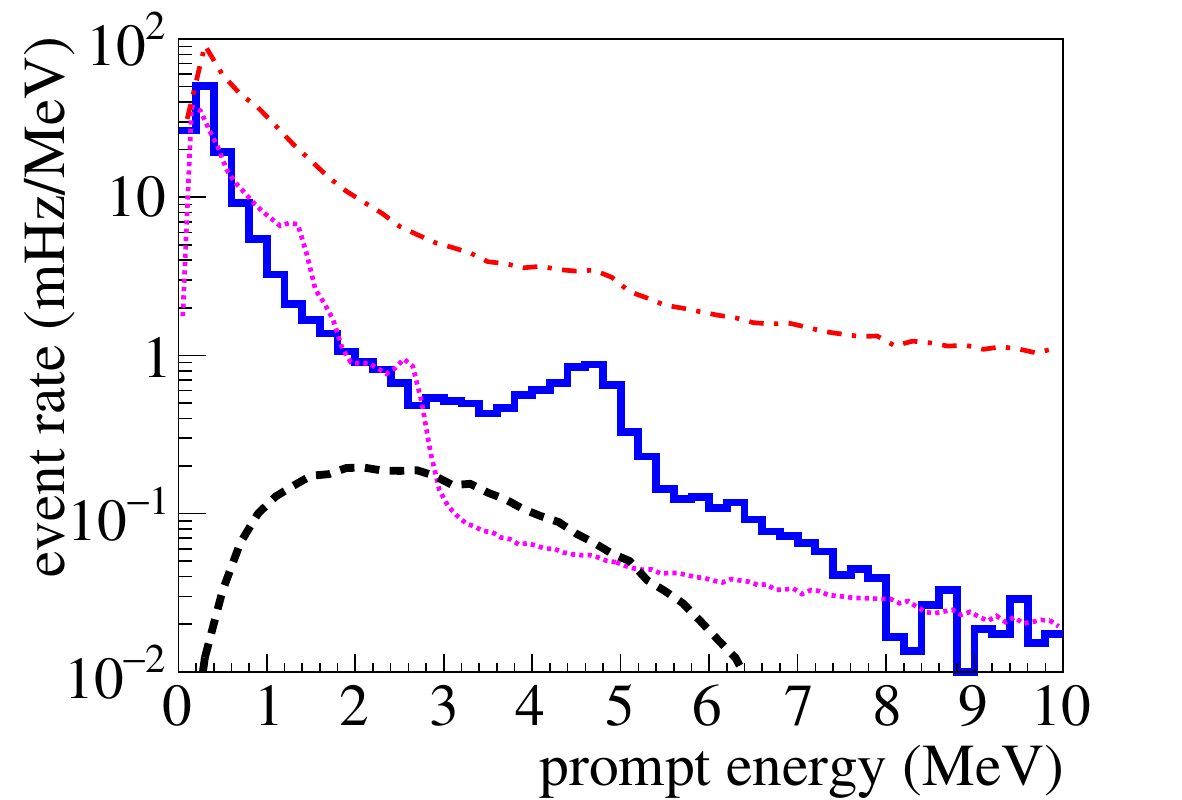} &
        \includegraphics[width=0.32\textwidth]{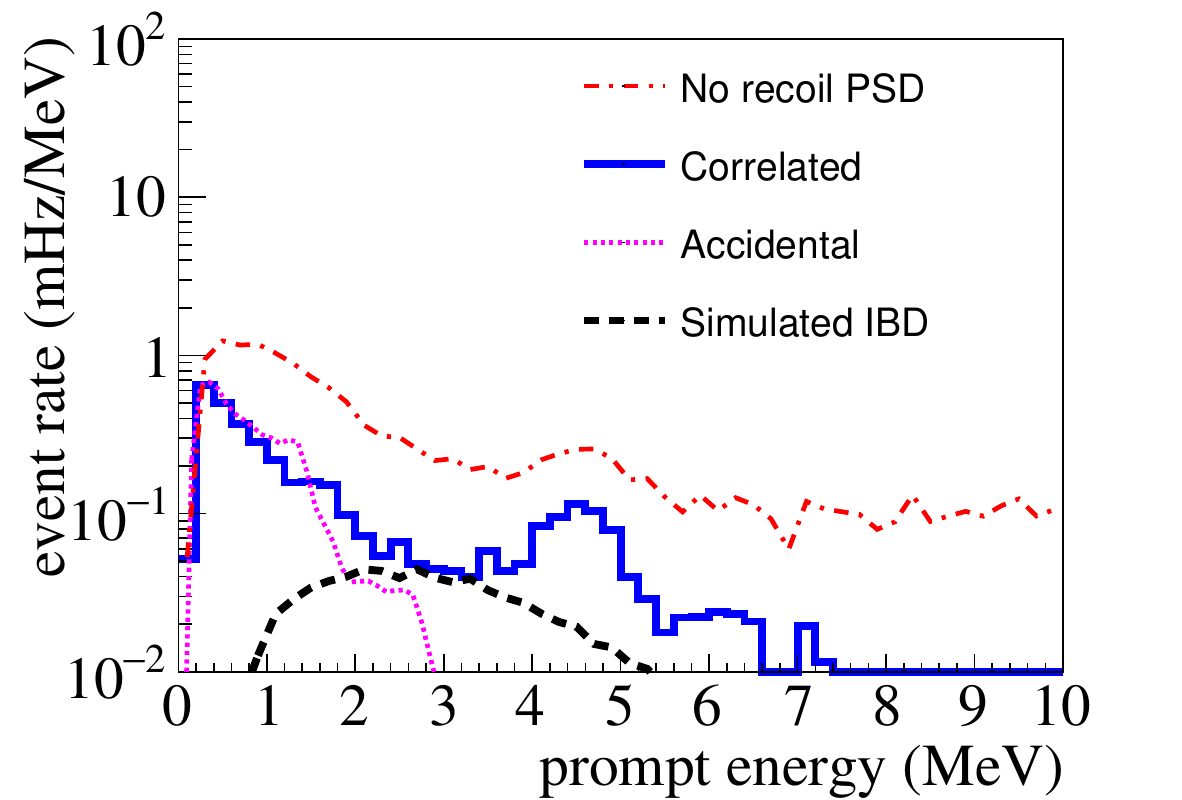} \\
    \end{tabular}
    \caption{
    IBD-like energy spectra for measured correlated (blue) and accidental (magenta) backgrounds, compared to the predicted IBD antineutrino  response for deployment 25~m from a 2900-MW$_{th}$ reactor (black). The predicted correlated backgrounds if the detector had no PSD sensitivity are shown (red). Different selection criteria:  
{\it (a)} Separation and PSD selections achieve good background rejection for a small $60$-kg detector. Topological selections based on {\it (b)} the pattern of energy depositions and {\it (c)} the relative geometry of hits in the prompt event can substantial improve $S:B$, albeit with a reduction of signal efficiency. 
    }
    \label{fig_IBDlike_data}
\end{figure*}

These selections are applied to background data collected by ROADSTR over a period of 43 days and the simulated IBD antineutrino event sample. 
To look for candidate IBD event pairs, a time-window scan of event clusters relative to neutron-capture-like events is performed.
Accidental backgrounds are subtracted using an off-time window. Energy spectra for simulated IBD, measured correlated backgrounds, and measured accidental backgrounds are shown in Fig.~\ref{fig_IBDlike_data}a. Daily event rates for the correlated signal and background event classes between $1.0$~MeV and $7.5$~MeV
are given in Table~\ref{tab_IBDlike}, as well as the expected signal-to-background ratio and an ``effective counts'' metric $S_{\text{eff}} = S^2/(S+2B)$,. Consider a counting experiment where $S+B$ and $B$ are measured independently, e.g. antineutrino measurements during reactor on and off periods.  
The $S_{\text{eff}}$ metric gives the number of background-free signal counts that would have equivalent statistical uncertainty after background subtraction for equal $S+B$ and $B$ integration periods. 

Also given in Table~\ref{tab_IBDlike} is the  efficiency for selecting IBD interactions (both prompt positron and delayed neutron capture),  $\varepsilon_{IBD}$.
From the IBD simulation, the fraction of IBD-generated neutrons that capture on $^6$Li within the detector is 68\%, with the remainder capturing on hydrogen or leaving the detector before capture.
The IBD efficiency values in Table~\ref{tab_IBDlike} indicate that the prompt and delayed selections reduce this fundamental ceiling on the IBD detection efficiency further.  
We note that a larger-scale implementation of this detection approach would increase the fraction of IBD neutrons that capture before leaving by decreasing the surface-to-volume ratio. For example, the macroscopic cross section ratio for captures on $^1$H and $^6$Li indicates an expected $^6$Li capture fraction of $\sim$85\% in an infinite detection volume, with the majority of capture happen within $\sim$10~cm of the production IBD vertex. 

Using the separation and PSD selections it is predicted that ROADSTR would achieve good performance for a $60$-kg prototype device at a large power reactor, with $S:B$ of almost $1:10$. With $S_{\text{eff}} = 2.8$~counts/day, a $3\sigma$ determination of a reactor On/Off transition could be achieved with $\sim 4$-days of signal and background data collection. 
Inclusion of the prompt PSD selection highlights the importance of this capability for rejection of background events involving a fast-neutron recoil: the $S:B$ and $S_{\text{eff}}$ metrics both increase by about an order of magnitude.

\begin{table}
\caption{Correlated IBD-like daily rates integrated between $[1.0,7.5]$~MeV passing several event selections for simulated IBD interactions ($S$) and  measured background ($B$). Also given are predictions for the background-free equivalent event rate ($S_{\text{eff}}$), signal-to-background ratio ($S:B$), and IBD detection efficiency $\varepsilon_{IBD}$.} 
\label{tab_IBDlike}
\centering
\begin{tabular}{|p{0.38\linewidth}|r|r|r|r|r|}
  \hline
  Selection            & $S$ & $B$  &  $S_{\text{eff}}$ & $S:B$ & $\varepsilon_{IBD}$ \\ \hline
  Separation only      & 70.4  & 9324 & 0.3  & 0.008 & 47.0\% \\
  Separation $+$ PSD   & 62.7  & 673  & 2.8  & 0.09 & 42.0\% \\
  $+$Energy topology   & 52.3  & 356  & 3.6  & 0.15 & 34.6\% \\
  $+$Geometry topology & 11.9  & 35 & 1.7  & 0.34 & 7.6\% \\ \hline
\end{tabular}
\end{table}

As noted above, for a surface-based deployment, the primary source of neutron-correlated backgrounds is cosmogenic fast neutrons. 
These can cause background due to processes that result in a neutron capture preceded by an event that passes the prompt PSD cut intended to select electromagnetic depositions. This can occur due to misidentification of proton recoils or the generation of $\gamma$ rays by processes like inelastic scatter on $^{12}\mathrm{C}$, where the excited nucleus returns to the ground state via emission of a $4.4$-MeV $\gamma$ ray,  
or the generation of multiple correlated neutrons where one captures on $^1$H followed by a capture on $^6$Li. 
The former process is clearly visible in the measured correlated background shown in Fig.~\ref{fig_IBDlike_data}.
Expected features due to naturally occurring radioactivity are also visible at 1.4~MeV (due to $\gamma$ rays from $^{40}$K) and 2.6~MeV (from $^{208}$Tl) in the accidental event spectrum.

Given the smaller segment size of ROADSTR relative to past 2D segmented prompt PSD-capable detectors like Bugey3 and PROSPECT, the ability to identify topological features associated with the annihilation of the IBD positron was also explored. 
The IBD positron will typically deposit its kinetic energy within centimeters of the IBD vertex and then annihilate with an electron producing two $\gamma$~rays with energy of $511$~keV travelling in opposite directions. 
These annihilation $\gamma$ rays have a mean free path of $\mathcal{O}$(10~cm) in organic scintillator and will deposit their energy via multiple Compton scatters. The $\sim$5-cm position resolution provided by the $x-y$ segmentation and $z$~position reconstruction of ROADSTR provides the opportunity to attempt identification of the spatial pattern of such IBD positrons which will, to some extent, be distinct from the IBD-like backgrounds with an electromagnetic prompt energy deposition (primarily capture of multiple correlated neutrons with the initial capture on $^1$H or inelastic scatter on $^{12}$C).

Selections are based on the pattern of ``hits'' (reconstructed position and energy) in the 2D~segment array. 
Multiple energy depositions within a bar, e.g. from multiple Compton scatters, cannot be resolved in ROADSTR, with a single energy and position value being inferred.  
Simulation data is processed via a weighted average to similarly return single energy, position, and PSD values~\cite{PROSPECT:2020sxr}.   

Two classes of topological selection have been investigated in this work. 
First, a selection is implemented to verify that the prompt event has energy depositions consistent with positron annihilation~\cite{Haghighat:2018mve}: a primary large energy deposition, with secondary depositions that sum to no more than $1022$~keV and that individually are no more than $511$~keV.
The second topological selection additionally considers the geometric relationship between the primary positron energy deposition and the secondary annihilation-$\gamma$-ray depositions, events with secondaries that are ``back-to-back'' about the primary. 

For the energy-based topological selection the hit with the largest energy  is identified as the primary positron deposition. If there are any hits in adjacent segments that are also adjacent in $z$, the largest of these is included in the primary position deposition. The energy of all other hits is then summed, with the event being rejected if this exceeds $1022$~keV. Finally, the energy of all other hits are examined, with the event being rejected if any exceeds $511$~keV. 
The results of this selection, applied in addition to the separation and PSD cuts listed above, are shown in Fig.~\ref{fig_IBDlike_data}b.

The geometry-based topological selection further examines the spatial relationship between the positron primary and pairs of secondary hits. An event is accepted if both of the following conditions are met for any pair of secondary hits:
\begin{enumerate}
\item  the perpendicular distance from the primary hit position to the line  drawn between the two secondary hit positions is less than one segment width (5.5\,cm). 
\item  the primary hit $(x,y,z)$ position falls within the axis-aligned box with corner vertices at the two secondary hit $(x,y,z)$ positions.
\end{enumerate}
The results of this selection, applied in addition to prior selections, are shown in Fig.~\ref{fig_IBDlike_data}c.

Both topology selections preferentially reduce background relative to signal, improving $S:B$ (Table~\ref{tab_IBDlike}). The energy-based topology selection has relatively modest impact on signal statistics, and the $S_{\text{eff}}$ metric improves so that a reactor  On/Off transition could be determined with $3$-days of signal and background data collection. The addition of the geometry-based selection improves $S:B$ to better than $1:4$, but reduces significantly signal efficiency and the $S_{\text{eff}}$ metric. 
This efficiency reduction results from the modest scale of the ROADSTR prototype relative to the  spatial extent of energy depositions from prompt positron annihilation.

The results for all selections are encouraging and can be expected to improve for larger detectors.
At larger scale, vetos tagging neutron-induced background via prompt and delayed PSD would be more effective, and as already noted neutron capture efficiency would improve. 
The  simple topological selections implemented here also have potential for improvement. 
In particular, the efficiency of selections based on geometry should improve with detector size as more energy from annihilation $\gamma$~rays
is contained and more sophisticated algorithms to identify those specific depositions are under development. 
Such selections would also benefit from full 3D segmentation~\cite{Haghighat:2018mve}.

\section{Summary and Impact on Future Nuclear Safeguards}

A 60-kg mobile antineutrino detector prototype has been constructed from $^6$Li-doped PSD plastic scintillator.  
The system has been used to collect calibration and background data since late 2022,  measuring correlated neutron-capture events analogous to IBD.
The demonstrated ability to calibrate the detector  without introducing radioactive sources is an important feature of the system.
Muons are used to set the energy scale, to calibrate the $z$-position sensitivity, and to synchronize the timing between different bars.
Natural cosmogenically-sourced neutrons are used to calibrate the neutron-capture signal and to determine the PSD cuts.
The $^{12}$B $\beta$ decay is used to provide point-like $z$-position sensitivity calibrations.
Bi-Po decay chain events, which provide a tagged $\alpha$~particle with an energy of 7.833~MeV, provide a direct data point for estimating scintillator quenching.

This is the first demonstration of $^6$Li-doped PSD plastic scintillator in the context of antineutrino reactor monitoring applications. 
By comparing measured background rates to a simulated IBD reactor signal, sufficient signal-to-background performance is predicted to observe a power reactor on-off transition within several days. 
This represents good performance relative to prior measurements by prototype systems of a similar scale. 
Simple topological selections taking advantage of the $5.5$-cm segmentation of ROADSTR also show promising results in rejecting IBD backgrounds.
When implemented at larger scale both positron identification and neutron capture efficiencies will improve, since the relevant $\mathcal{O}$(10~cm) length scale for $\gamma$~ray and neutron transport is comparable to the dimensions of this prototype system.   
The technology described here will form the basis for a detector system approximately 5~times larger as part of the Mobile Antineutrino Demonstrator Project in the near future.
Studies using the ROADSTR detector to characterize the effectiveness of varied shielding configurations and backgrounds in locations relevant to monitoring applications are also underway.  

\section{Acknowledgements}
This work was performed under the auspices of the U.S. Department of Energy by Lawrence Livermore National Laboratory under Contract DE-AC52-07NA27344. This work was supported by the LLNL-LDRD Program under Project No. 20-SI-003 and by the U.S. Department of Energy Office of Defense Nuclear Nonproliferation Research and Development. LLNL-JRNL-2005361.

The authors acknowledge contributions from Oluwatomi Akindele, Phillip Hamilton, Andrew Mabe, and Nathan Eric Robertson at the early stages of this project. 

\section{Data Availability}
Data available from the authors upon reasonable request.

\newpage
\bibliography{ref}

\end{document}